\documentclass[journal]{IEEEtran}

\usepackage{color}

\usepackage{graphicx}
\usepackage{algorithm}
\usepackage[noend]{algpseudocode}
\usepackage{amsmath}
\usepackage{amssymb}
\usepackage[export]{adjustbox}
\usepackage{subfigure}
\usepackage{float}
\usepackage[flushleft]{threeparttable}
\usepackage{multirow}
\usepackage[switch]{lineno}
\usepackage{array}

\ifCLASSINFOpdf
 
\else

\fi

\begin{document}

\title{Learning in Memristive Neural Network Architectures using Analog Backpropagation Circuits}


\author{Olga Krestinskaya,~\IEEEmembership{Graduate Student Member, IEEE,}
Khaled Nabil Salama,~\IEEEmembership{Senior Member,~IEEE} 
	  and Alex Pappachen James,~\IEEEmembership{Senior Member,~IEEE}  
\thanks{O. Krestinskaya is a graduate student and research assistant with the Bioinspired Microelectronics Systems Group, Nazarbayev University}
\thanks{K. N. Salama is a Professor in King Abdullah University of Science and Technology in Kingdom of Saudi Arabia and principal investigator of Sensors Lab.}
\thanks{A.P. James is the Chair of Electrical and Computer engineering with the School of Engineering, Nazarbayev University, e-mail: apj@ieee.org.}%
}

\maketitle

\begin{abstract}
The on-chip implementation of learning algorithms would speed-up the training of neural networks in crossbar arrays. 
The circuit level design and implementation of backpropagation algorithm using gradient descent operation for neural network architectures is an open problem. In this paper, we proposed the analog backpropagation learning circuits for various memristive learning architectures, such as Deep Neural Network (DNN), Binary Neural Network (BNN), Multiple Neural Network (MNN), Hierarchical Temporal Memory (HTM) and  Long-Short Term Memory (LSTM). The circuit design and verification is done using TSMC 180nm CMOS process models, and TiO$_2$ based memristor models. The application level validations of the system are done using XOR problem, MNIST character and Yale face image databases.

\end{abstract}

\begin{IEEEkeywords}

Analog circuits, Backpropagation, Learning, Crossbar, Memristor, Hierarchical Temporal Memory, Long-Short Term Memory, Deep Neural Network, Binary Neural Network, Multiple Neural Network
 
\end{IEEEkeywords}

\IEEEpeerreviewmaketitle

\section{Introduction}

\IEEEPARstart{T}{he}  developments in Internet of Things (IoT) applications led to the demand to develop the near-sensor edge computation architectures \cite{shi2016edge}. The edge computing provides motivation to develop near-sensor data analysis that support non-Von Neumann computing architectures such as neuromorphic computing architectures \cite{vermesan2017internet,hoang2015will}. In such architectures, implementing the on-chip neural network learning remain as an important task that determines its overall effectiveness and use.




{There are several works proposing the implementation of memristive neural network with backpropagation algorithm in digital and mixed-signal domain domain \cite{8030206,8060399,burr2017neuromorphic,nn3,7966300,kim2017analog,tsai2018recent}. However, the analog learning circuits based on conventional backpropagation learning algorithm \cite{zhang2017memristor,negrov2017approximate,7966300,nn1,7867846} in memristive crossbars have not been fully implemented. } The implementation of such learning algorithm opens up an opportunity to create an analog hardware-based learning architecture. This would transfer the learning algorithms from the separate software and FPGA-based units to on-chip analog learning circuits, which can simplify and speed up the learning process.


Extending our previous work on analog circuits for implementing backpropagation learning algorithm \cite{ISCASs}, we present a system level integration of the analog learning circuits with that of traditional neuro-memristive crossbar array. We illustrate how this learning circuit can be used in different biologically inspired learning architectures, such as three layer Artificial Neural Networks (ANN), Deep Neural Networks (DNN) \cite{8060399, 7966300}, Binary Neural Network (BNN) \cite{8297384}, Multiple Neural Network (MNN) \cite{363444}, Hierarchical Temporal Memory (HTM) \cite{25} and Long-Short Term Memory (LSTM) \cite{6795963}. 

{The algebraic and integro-differential operations of backprogation learning algorithm, which are difficult to accurately implement on a digital system, are available inherently on the analog computing system. Further, modern edge-AI computing
solutions warrant intelligent data processing at sensor levels, and analog system can reduce the demands for having high speed
data converters and interface circuits. The proposed analog backpropagation learning circuit 
enables a natural on-chip analog neural network architecture implementation which is beneficial in terms of processing speed, reducing overall power and lesser complexity, comparing to digital counterparts.
}

{
The main contributions of this paper include the following.
\begin{itemize}
\item We introduce the complete design of analog backpropagation learning circuit proposed in \cite{ISCASs} with control switches sign control circuit and weight update unit.
\item We illustrate how the proposed backpropagation learning circuit can be integrated into different neuromorphic architectures, like DNN, BNN, MNN, LSTM and HTM.
\item We show the implementation of additional activation functions that are useful for various neuromorphic architectures.
\item We verified the proposed architecture for XOR problem, handwritten digits recognition and face recognition, and illustrate the effect of non-ideal behavior of memristors on the performance of the system.
\end{itemize}
}

{
This paper is organized into following sections:
Section \ref{s2} introduces the relevant background of the learning architectures and backpropagation algorithm. Section \ref{s3} describes the proposed architecture of the backpropagation and the circuit level design. Section \ref{s4} illustrates how the proposed backpropagation circuits can be integrated into different learning architectures. Section \ref{s5} contains the circuit and system level simulation results. Section \ref{s6} discusses advantages and limitations of the proposed circuits and introduces the aspects of the design that should be investigated in future, and Section \ref{s7} concludes the paper. There is also a supplementary Material that includes the expanded background information, explicit explanation of the proposed circuit, the device parameters of the main backpropagation circuit proposed in Section \ref{s3} and simulation results for the learning circuit performance.
}

\section{Background}
\label{s2}

\subsection{Learning algorithms and biologically inspired learning architectures}

Three main brain inspired learning architectures that we consider in this work are neural networks \cite{8060399, 8297384, 363444,krestinskaya2018neuro}, HTM \cite{25} and LSTM \cite{6795963}.

\subsubsection{Neural Networks} 

There is a variety of the architectures and learning algorithms for the neural networks. In this work, we integrate analog learning circuits to three different types of artificial neural network \cite{jain1996artificial}: DNN \cite{8060399, 7966300}, BNN \cite{8297384} and MNN \cite{363444}. DNNs consist of many hidden layers and can have various combination of activation functions between the layers. Deep learning neural networks are useful for classification \cite{els1}, regression \cite{6932438}, clustering \cite{7010902} and prediction tasks \cite{8290807}.  A neural network which uses any combination of binary weight or hard threshold activation functions is typically known as BNN. There have been several successful implementations of BNN algorithms in software \cite{557699, 125859, 6797577,5260357} and an attempt to implement it on hardware \cite{secco2018supervised, 8052915,bnn}. The analog hardware implementation of BNN system with learning remains as an open problem \cite{hubara2016binarized,bnn}. MNN is a systematic arrangement of the artificial neural networks that can process information from different data sources to perform data fusion and classification. Multiple neural network implies that the data from different data sources, such as various sensors, is applied to separate neural networks, and the output from each network is fetched into the decision network. This approach allows simplifying the complex computation processes, especially when there is a large number of data sources \cite{363444}. The analog hardware implementation of MNN is another new idea proposed in this paper.

\subsubsection{Hierarchical Temporal Memory} 

HTM is a neuromorphic machine learning algorithm that mimics the information processing mechanism of neocortex in the human brain. HTM architecture is hierarchical and modular, and it enables sparse processing of information. HTM is divided into two parts: (1) Spatial Pooler (SP) and (2) Temporal Memory (TM) \cite{hawkinsintelligence,krestinskaya2018hierarchical}. The main purpose of the HTM SP is to encode the input data and produce its sparse distributed representation that finds application in various data classification problems. The HTM TM is primarily known for contextual analysis, sequence learning \cite{7727380} and prediction tasks \cite{bami2016, george2005hierarchical}. The HTM SP consists of four main phases: (1) initialization, (2) overlap, (3) inhibition, and (4) learning. There are several hardware implementations proposed for the HTM SP, such as conventional HTM SP \cite{fedorova2016htm} and modified HTM SP \cite{tcad}. Both architectures are based on memristive devices located in the initialization and overlap stages. The hardware implementation of the learning stage for HTM SP has not been proposed yet. According to \cite{25}, the backpropagation algorithm can be one of the approaches to updating weight in the HTM SP.

\subsubsection{Long-Short Term Memory} 

LSTM is a cognitive architecture that is based on the sequential learning and temporal memory processing mechanisms in the human brain. LSTM processing relies on state change and time dependency of processed events. The LSTM algorithm is a modification of recurrent neural network that takes into account history of processed data and controls the information flow through gates \cite{olstm3,7508408}. LSTM is used in a wide range of applications in the contextual data processing based on prediction making and natural language processing. Hardware implementation of LSTM is a new topic studied in \cite{kam,smagulova2018design}.

\subsection{Backpropagation with gradient descent}

In this paper, an analog implementation of gradient descent backpropagation algorithm \cite{chauvin1995backpropagation,mehrotra1997elements} is proposed for different neural network configurations. 
The algorithm consists of four steps: forward propagation, backpropagation to the output layer, backpropagation to the hidden layer, and weight updating process. In this section, we present the main equations of backpropagation algorithm with gradient descent for a three-layer neural network with sigmoid activation function to relate it to the proposed hardware implementation.

In the forward propagation step, the dot product of input matrix $X$ and weighted connections between input layer and hidden layer $w_{12}$ is calculated and passed through the sigmoid activation function: $Y_h=\sigma(X\cdot w_{12})$, where $Y_h$ is an output of the hidden layer \cite{7458870,8030932}. The forward propagation step is repeated in all the neural network layers. The output of the three-layer network $Y_o$ is calculated as $Y_o=\sigma(Y_h\cdot w_{23})$, where $w_{23}$ is the matrix representing the weighted connections between the hidden and output layers. 

The backpropagation algorithm uses the cost function defined in Eq.\ref{eq3} for the calculation of derivative of error with respect to the change in weight. In Eq.\ref{eq3}, $E$ is an error, $N$ is a number of neurons in the layer, $y_{target}$ is an ideal output and $y_{real}$ is the obtained output after the forward propagation \cite{leonard1990improvement}.

\begin{equation}
E=\frac{1}{2}\sum_{i=1}^{N} (y_{target}-y_{real})^2
\label{eq3}
\end{equation}

Equation \ref{eq4} shows the calculation of the derivative of output layer error $E_o$, where $\delta$ denotes the rate of change of the error with respect to the weight $w_{23}$ \cite{7458870,8030932}. The error for the output layer $e$ is calculated as a difference between the expected neural network output $Y$ and real output of the network $Y_o$: $e=Y-Y_o$. The derivative of the sigmoid function is the following: $\frac{\partial Y_o}{\partial w_{23}} = Y_o(1-Y_o)$.

\begin{equation}
\frac{\partial E_o}{\partial w_{23}}=Y_h\cdot \delta _2 = Y_h \cdot (e \odot \frac{\partial Y_o}{\partial w_{23}})
\label{eq4}
\end{equation}

The derivative of error for the hidden layer is shown in Eq. \ref{eq5}, where $X'$ is an inverted input matrix and $e_h$ is the error of the hidden layer. The error $e_h$ is calculated propagating back $\delta_2$ as following: $e_h=\delta _2 \cdot w_{23}'$. And the derivative of the hidden layer output $E_h$ is the same as in the output layer: $\frac{\partial Y_h}{\partial w_{12}} = Y_h(1-Y_h)$ \cite{golden1996mathematical}.

\begin{equation}
\frac{\partial E_h}{\partial w_{12}}=X'\cdot \delta _1 = X' \cdot (e_h \odot \frac{\partial Y_h}{\partial w_{12}})
\label{eq5}
\end{equation}

In the final stage, the weight update calculation is performed using Eq. \ref{ee1} and Eq. \ref{ee2}, where $\eta$ is the learning rate responsible for the speed of convergence. The optimized learning rate depends on the type and number of inputs and number of the neurons in the hidden layers \cite{8030932, golden1996mathematical}.

{
\begin{equation}
\Delta w_{23} =\frac{\partial E_o}{\partial w_{23}} \times \eta 
\label{ee1}
\end{equation}
\begin{equation}
\Delta w_{12} =\frac{\partial E_h}{\partial w_{12}} \times \eta
\label{ee2}
\end{equation}
}

The weight matrices are updated considering the calculated change in weight: $w_{23\_new}=w_{23}+\Delta w_{23}$ and $w_{12\_new}=w_{21}+\Delta w_{12}$.

\section{Backpropagation with memristive circuits}

\label{s3}
\subsection{Overall architecture}

This subsection illustrates the overall implementation of the backpropagation algorithm on hardware, while the details of the implementation of the activation functions and particular blocks are shown in Section \ref{sec3b}.

\begin{figure}
\centering
\includegraphics[width=90mm]{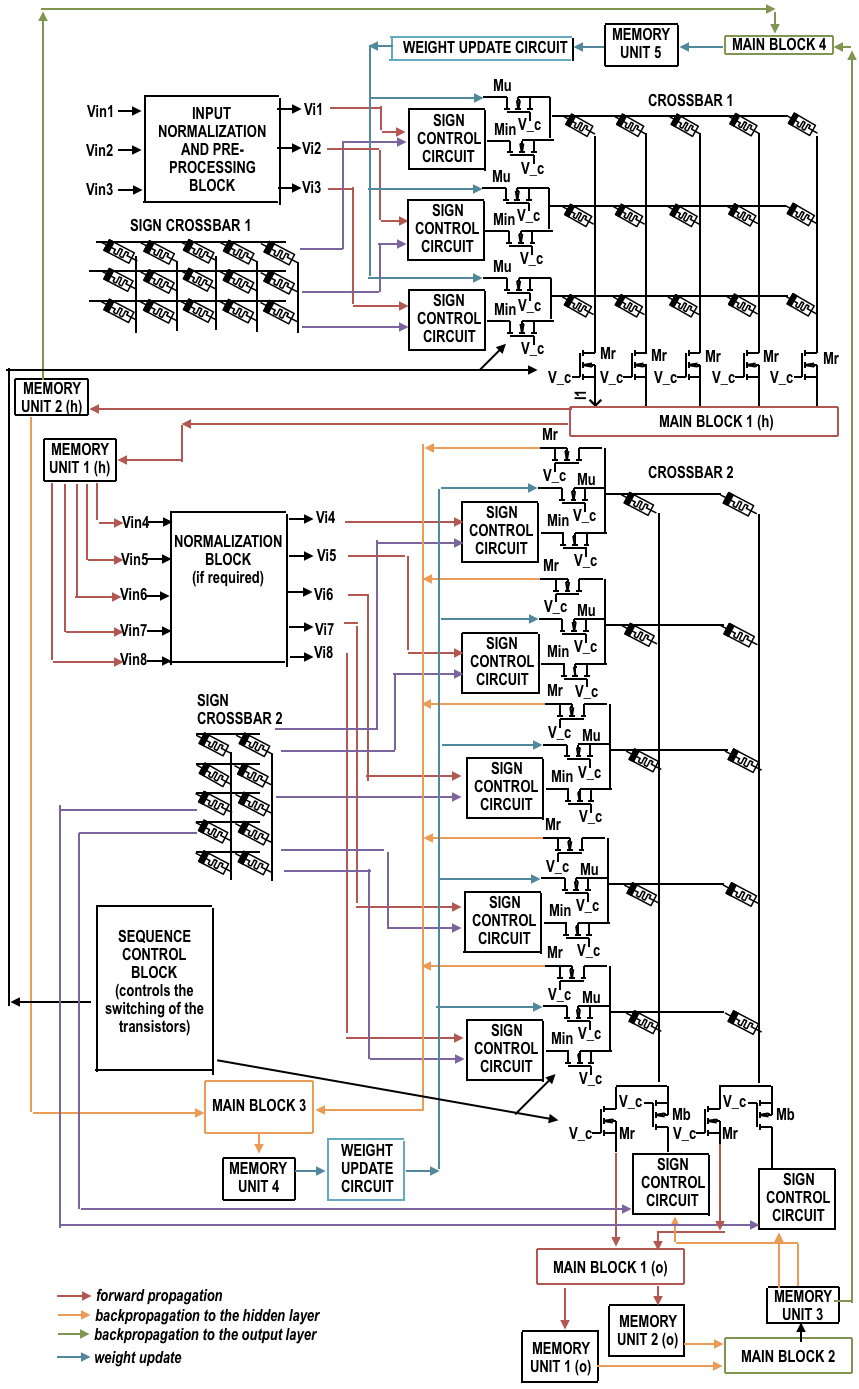}
\caption{Overall architecture of the proposed analog backpropagation learning circuits for memristive crossbar neural networks. In the forward propagation process, MAIN BLOCK 1 (MB1) is involved. The backpropagation through the output layer is performed by MAIN BLOCK 2 (MB2) and MAIN BLOCK 4 (MB4). The backpropagation through the hidden layer is performed by MAIN BLOCK 3 (MB3). The weight update process of the output layer and the hidden layer is performed by MB4 and MB3, respectively. {The blocks with the notation (o) correspond to output layer and the block with notation (h) correspond to hidden layer.}}
\label{f1}
\end{figure}

The proposed hardware implementation of the learning algorithm is illustrated in Fig. \ref{f1}.   {Depending on the application requirements and limitations of the memristive devices, the inputs to the system can be either binary or non-binary. For example, for the HTM applications, the inputs are non-binary \cite{tcad}, whereas, for the binary neural network, inputs can be binary \cite{hubara2016binarized}. The outputs of the neural network can also be binary or non-binary, depending on the activation function. 
Memristive crossbar arrays emulate the set of synapses between the neurons in the neural network layers. The synapses can also be binary or non-binary depending on the applications and practical limitations of programming states of memristor device. } While an ideal non-volatile memristor device can store and be programmed to any particular value between $R_{ON}$ and $R_{OFF}$, the real memristor devices can have problems with switching to the intermediate resistive values. It is easier and simpler to switch the memristor to either $R_{ON}$ and $R_{OFF}$ state. The implementation of the analog weights is also possible using 16-level $\textrm{Ge}_2\textrm{Sb}_2\textrm{Te}_5$ (GST) memristors \cite{xiao2017gst}. However, the memristor technology is not mature like CMOS, and even if the memristor can be precisely programmed and work accurately under the controlled environment in the lab, the behavior of the memristor in the multi-level large-scale simulation still needed to be verified.
Therefore, the binary synapses are the easiest to be implemented. 

{
The example shown in Fig. \ref{f1} demonstrates the basic three-layer ANN with the proposed backpropagation architecture, control circuit and weight update circuits. The neural network has three input neurons, two output neurons and five neurons in a hidden layer. The operation of the crossbar and switching between forward propagation, backpropagation and weight update operations is controlled by the switching transistors $M_{in}$, $M_{u}$ and $M_{r}$, which in turn are controlled by the sequence control block. The CROSSBAR 1 corresponds to the set of synaptic connections between the input layer and the hidden layer, and the CROSSBAR 2 represents the synapses connecting the hidden layer to the output layer. The three input signals are shown as $V_{in1}$, $V_{in2}$  and $V_{in3}$, and the corresponding normalized input signals are shown as $V_{i1}$, $V_{i2}$  and $V_{i3}$, respectively. The range of output signals from the normalization circuit depends on the application, limitations of memristors and linearity of the switch transistors. The inputs are fetched to the rows of the CROSSBAR 1 i.e. to the input switching transistors $M_{in}$.} Each memristor in a single column of the crossbar corresponds to the connections of all inputs to a particular single output. The crossbar performs dot product multiplication of inputs and the weights of a single column. The output of the multiplication for feed-forward propagation is read from the NMOS read transistor $M_{r}$ connected to a crossbar column. In this work, we investigate the approach, when the output of the crossbar is represented by a current flowing through the read transistor. However, the configuration can be changed to the voltage output from the crossbar column if required. The voltage-based approach is more complicated than current-based approach because the amplifier or OpAmp-based buffer is required to read the voltages without the loading effect from the following interfacing circuits. The current-based approach requires only the use of a current mirror ensuring the reduction of the on-chip area, {and also compatible with simple current driven sigmoid implementation}.

{
The parameters of input transistors $M_{in}$, read transistors $W_{r}$ and corresponding control signals $V_c$ should be selected carefully, considering the range of input signal. When the signals are propagated back through in the CROSSBAR 2, the propagated error can be both positive and negative. Therefore, if it is important to make sure that the size of the transistors and $V_c$ are set to eliminate the current  when transistor in OFF state and conduct the current in linear region when transistor is ON. These parameters should be adjusted depending on the technology.}

{
The outputs from the crossbar columns are read sequentially one at a time to avoid the interference with the currents from the other columns. The read-transistors $M_{r}$ at the end of each column are used to switch ON and OFF the columns and maintain the order of the reading sequence for the forward propagation process.   
As the negative resistance is not practical to implement in the memristive device, the negative weights are implemented using either input controller, which changes the input sign according to the sign of the synapse weight, or by the sign control circuit and additional crossbar that stores the sign of the weights. Fig. \ref{f1} illustrates the approach with sign control circuit and additional SIGN CROSSBAR 1 and SIGN CROSSBAR 2.
}

{
MAIN BLOCK 1 (MB1) in Fig. \ref{f1} performs the forward propagation and is responsible for the calculation of the activation function. MB1 (h) and MB1 (o) correspond to hidden layer and output layer, respectively. Depending on the application, the activation function can include the sigmoid, derivative of the sigmoid, tangent, derivative of the tangent, approximate sigmoid and approximate tangent functions. The approximate functions represented here refer to "hard" logical threshold sigmoid and tangent as shown in \cite{golden1996mathematical}. To implement the conventional backpropagation algorithm with gradient descent, MB1 performs the calculation of the sigmoid and the sigmoid derivative functions. The outputs of MB1 are stored in MEMORY UNIT (MU) 1(h) and used as the inputs to CROSSBAR 2. The outputs of MU1(h) can be normalized by normalization circuit. The output currents from the second crossbar are fetched to the second MB1 and the final output of the feed-forward propagation are obtained from MB1(o) and stored in MU1(o) for further application in backpropagation. The final outputs depend on the activation function. In addition, MB1 produces the outputs of the activation function, stored in MU1(h) and MU1(o), and derivative of the activation function, stored in MU2(h) and MU2(o) that are useful for backpropagation process.
}

{
After the forward propagation process, the backpropagation process is implemented.  The sequence control block switches off the column read transistors $M_r$ of both crossbars and switches on the row read transistors $V_r$ of CROSSBAR 2.
The column transistors $M_{in}$ of CROSSBAR 2 are switched ON to propagate the inputs MU3, which stores the outputs of MAIN BLOCK 2 (MB2) corresponding to the propagation through the output layer. It ensures that the propagation is performed in the opposite direction. If the neural network contains more than three layers, all crossbars except the first crossbar corresponding to the synaptic weights between the input and the first hidden layer are reconnected to perform backpropagation operation. The backpropagation process through the output layer is implemented using MB2 and MAIN BLOCK 4 (MB4), and the backpropagation through the hidden layer corresponds to the MAIN BLOCK 3 (MB3). The possible architecture of analog memory unit is illustrated in \cite{tcad,aidana,aidanaconf}. 
}

The final stage in the backpropagation algorithm is the weight update stage, where the values of the memristors are updated based on the specific rules. As the crossbar values are read and processed sequentially, MEMORY UNIT 4 and 5 store the update value before the update process starts. The weight update process is implemented by applying the voltage pulse of a particular duration and amplitude across each memristor. {The update pulse depends on the required change of the weights, calculated gradient of error, and the memristor type and technology. The amplitude of the update pulse depends on the outputs of MB2 and MB4 and calculated by weight update circuit. While the duration of the pulses are controlled by the sequence control circuit.
The update process of the memristor is controlled by transistors $M_{in}$ and $M_{r}$. To update the memristor, corresponding row transistors $M_{in}$ and column transistor $M_r$ are switched ON. 
As the main architecture for the synapses that we consider in this work is 1M, each memristor is updated either one at a time. However, this process is slow, and in particular cases memristor weight can be updated in several cycles as illustrated in \cite{cam}. If the two state memristors are used in the crossbar ($R_{ON}$ and $R_{OFF}$, which is useful for binary neural network), the update process can be performed in two cycles: (1) update of all $R_{ON}$ memristors, which should be switched to $R_{OFF}$ and (2) update of all $R_{OFF}$ memristors, which should be switched to $R_{ON}$. This method is useful, when the other states between $R_{ON}$ and $R_{OFF}$ are not important for processing. Such method increases the speed of update process, however it is useful for neural network with binary synapses.}

\subsection{Circuit-level implementation of main blocks in backpropagation algorithm}
\label{sec3b}
\begin{figure*}[h]
\centering
\includegraphics[width=140mm]{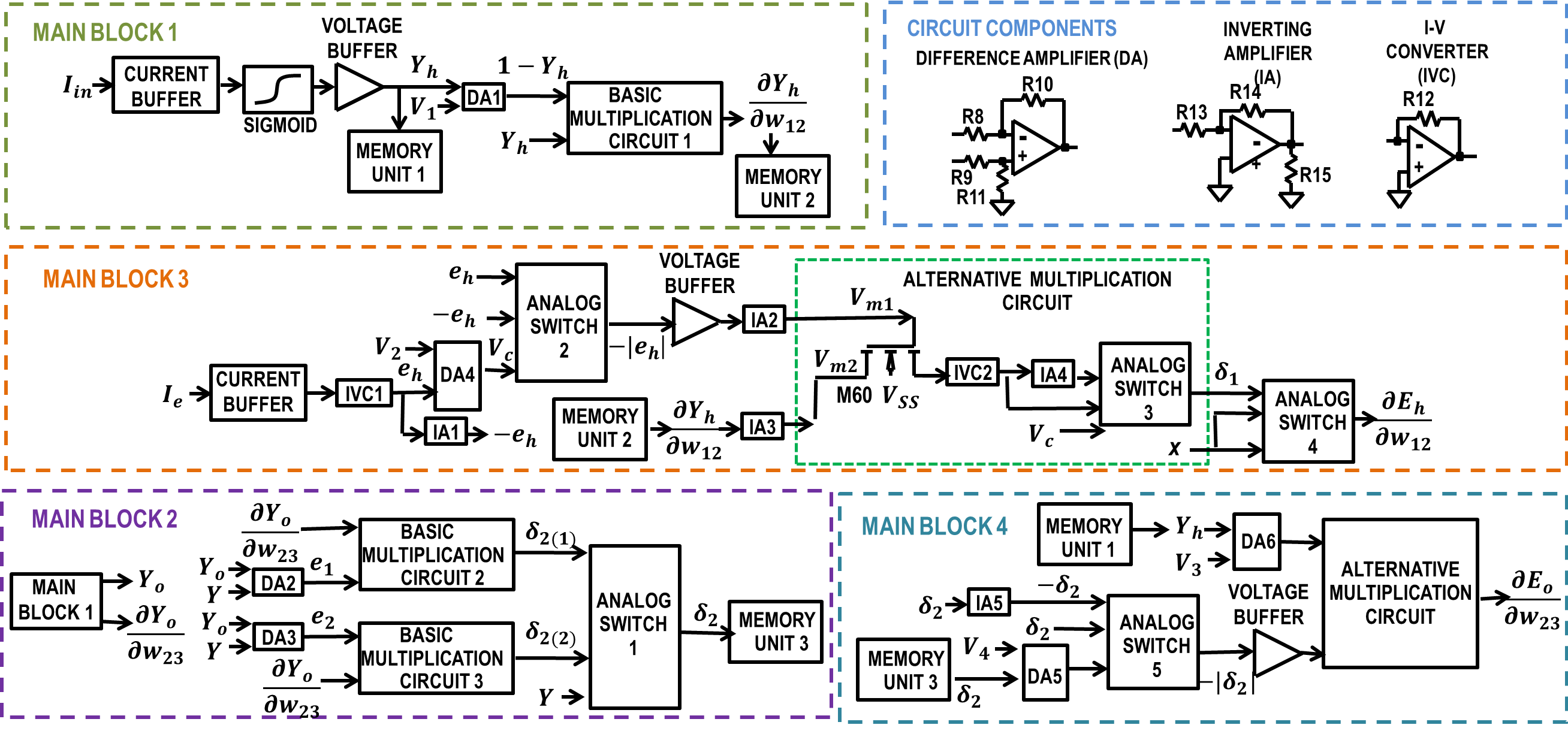}
\caption{The circuit level architecture of the proposed backpropagation implementation. The separate implementation of the MB1, MB2, MB3 and MB4 is illustrated. In addition, the involved circuit components, such as DA, IA and IVC, are shown.}
\label{f2}
\end{figure*}

\begin{figure*}[h]
\centering
\includegraphics[width=180mm]{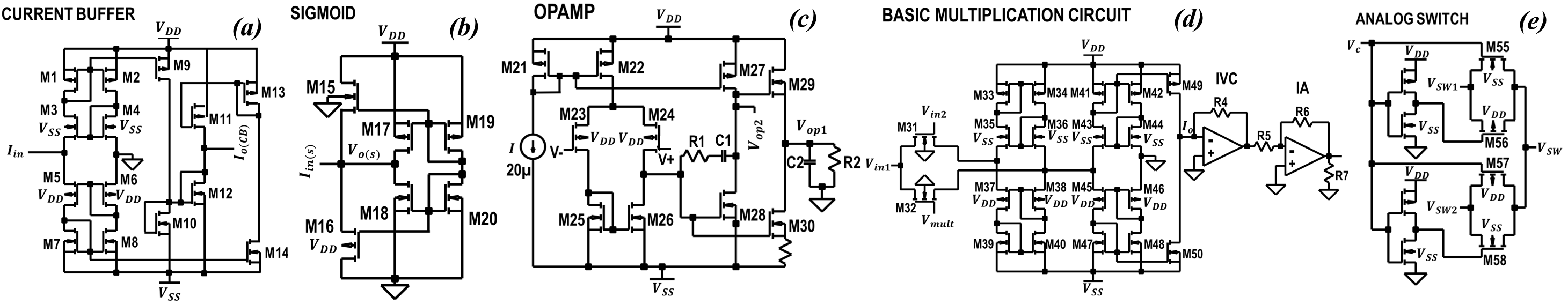}
\caption{Circuit components: (a) The current buffer circuit which is connected to the read transistor $M_r$ in Fig. \ref{f1}. The circuit is used in MB1 and MB3 to eliminate the loading effect of the activation function to the performance of the crossbar. (b) Sigmoid activation function used in MB1 inspired from \cite{5719144}. (c) Two stage OpAmp design used for all OpAmp based components in the proposed analog memristive learning circuit. (d) Multiplication circuit based on the Hilbert multiplier principle. The circuit is used in MB1 and MB2. (e) Analog switch design used in MB2, MB3 and MB4.}
\label{cnew}
\end{figure*}

{
This section briefly introduces the proposed architecture for backpropagation circuits shown in \cite{ISCASs}, while the detailed explanation of the circuit and all circuit parameters are provided in Supplementary Material (Section III).
In this work, the analog circuits for the proposed backpropagation implementation are designed for 180nm CMOS process.
The circuit level implementation of all backpropagation blocks is illustrated in Fig. \ref{f2}, while the components used in the circuits are shown in Fig. \ref{cnew}.
 MB1 performs the forward propagation for the conventional backpropagation architecture. MB2 performs the backpropagation process through the output layer, which is finished by MB4. MB3 performs the backpropagation through the hidden layer.
The circuit level implementations of components from the main backpropagation blocks are shown in Fig. \ref{cnew}. Fig. \ref{cnew}(a)
 illustrates the implementation of current buffer. Fig. \ref{cnew}(b) shown the implementation of adjusted sigmoid function proposed in  \cite{5719144}. Fig. \ref{cnew}(c) shows the OpAmp circuit. Fig. \ref{cnew}  (d) illustrated the multiplication circuit based on the current difference in transistors $M_{31}$ and $M_{32}$. Fig. \ref{cnew} demonstrated the implementation of analog switch circuit.
}


{
\subsection{Sign control circuit}
}

\label{sec3c}

\begin{figure}
\centering
\includegraphics[width=60mm]{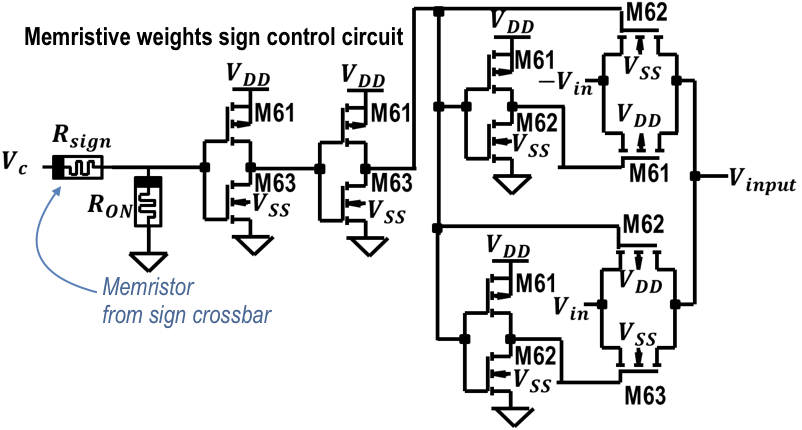}
\caption{Memristive weight sign control circuit that can be integrated to the crossbar to control the weight of the synapse or applied as an external circuit with a separate memristors to store the sign of the weight.}
\label{f8}
\end{figure}

As the neural network weights can be both positive and negative, and the negative weight cannot be practically implemented by the memristor, the implementation of the additional weight control circuit is required. For each negative weight, the sign of the input voltage is changed. There are two possible ways to implement the sign. One of the solutions is to store the sign for each sequence in the external storage unit and apply it to the circuit with the weight normalization circuit. The other solution is to store the sign of each weight in the additional memristive crossbar elements. A memristive weight sign control circuit shown in Fig. \ref{f8} is proposed. 

The sign of each memristor in the crossbar is stored in the memristive crossbar or separate memristors as $R_{ON}$ or $R_{OFF}$. The analog sign read circuit follows the memristor storing the sign of the weight. There are two possible solutions. The first is to implement a single analog sign read circuit and switch it between the memristors in the crossbar, which requires additional on-chip area and storage. And the second and more effective solution is to implement the number of sign read circuits equivalent to the number of rows in a crossbar, which allows reading the sign of all the memristors in a single column. This allows achieving the trade-off between the required area, power and processing time. In Fig. \ref{f8}, the sign of the memristor representing the weight is stored in the memristor $R_{sign}$. When the sign is read, $V_{c}=1.25V$ is applied. If $R_{sign}$ is set to $R_{ON}$, the output of the analog switch $V_{sign}$ is positive and vice versa. The weight sign read circuit acts as a switch. If $R_{sign}=R_{ON}$, the voltage $V_{c}$ is above the switching threshold, it selects, and outputs the positive voltage $V_{in}$, which is the input voltage to the crossbar. If $R_{sign}=R_{OFF}$, the voltage drops and $V_{c}$ is below the switching threshold, and the switch outputs the voltage $-V_{in}$. The parameters of the transistors are the following: $M_{61}=0.18\mu/0.72\mu$,  $M_{62}=0.18\mu/10.36\mu$ and  $M_{63}=0.18\mu/0.36\mu$. The transistors in the circuit have an underdrive voltage $V_{DD}=1V$.


Comparing to the existing implementations of negative voltages in the crossbar array \cite{hu2016dot,truong2014new,hu2012hardware}, the method for sign control reduces the complexity of the implementation and ensures the stability of the output. For example, the crossbar in \cite{truong2014new} can perform dot product multiplication for both positive and negative signals. However, the system is complex because of the amplifiers that perform subtraction of the voltages. To ensure the amplification is not affected by the following circuits, such amplifier should include the capacitor, which increases the on-chip area of the circuit.
Also, this way of implementing negative signals require the accuracy preprocessing stage and additional adjustment of the input signals. A similar method of implementing negative sign is shown in \cite{hu2012hardware}. It involves a set of summing amplifiers with resistors, which also consume a large amount of area. 

\begin{figure}
\centering
\includegraphics[width=40mm]{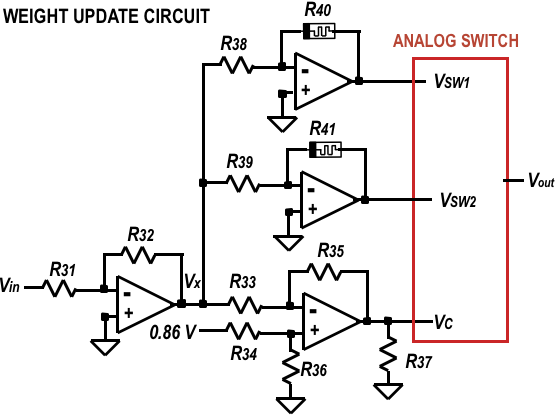}
\caption{{Memristive weight update circuit that converts the weight update value from MB3 and MB4 to the pulse of particular amplitude. The used analog switch is shown in Fig. \ref{cnew} (e).}}
\label{wuc}
\end{figure}

{
\subsection{Weight update circuit}
The implementation of memristive weight update circuit is illustrated in Fig. \ref{wuc}. The weight update circuit determined the pulse amplitude required to program the memristor in a crossbar array, depending on the calculated weight update value by MB2 and MB4.
The circuit is adaptable for different learning the due to the application of memristive devices $R_{40}$ and $R_{41}$ in the amplifiers. All the resistors in weight update circuit are $1k\Omega$ and the memristors are programmed considering the required learning rate. As the negative (to switch from $R_{OFF}$ to $R_{ON}$) and positive (to switch from $R_{ON}$ to $R_{OFF}$) programming amplitudes are not of the same amplitude for memristive devices, the analog switch selects the amplitude of the signal based on sign of the input voltage from MB1 and MB4. The implementation of analog switch is shown in Fig. \ref{cnew} (e) The shifted input signal is applied to analog switch control  $V_c$ that determines, which input to the switch should be selected $V_{SW1}$ or $V_{SW2}$. The input to $V_{SW1}$ corresponds to the positive input voltage, while $V_{SW2}$ corresponds to the negative input voltage. 
}

\begin{figure}
\centering
\includegraphics[width=90mm]{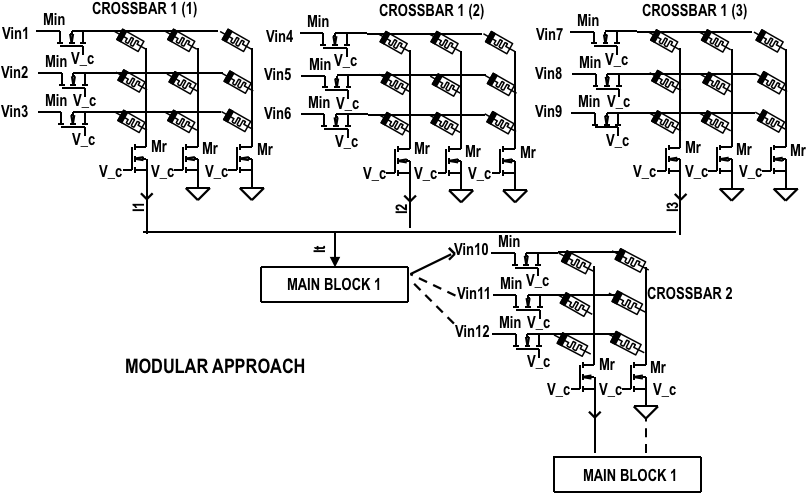}
\caption{{Modular approach to reduce the leakage current and complexities in programming of 1M array.}}
\label{mod}
\end{figure}

{
\subsection{Modular approach}
As large scale crossbars usually suffer from leakage currents, the most widely used architecture for the crossbar synapses is 1 transistor 1 memristor $1T1M$ \cite{krestinskaya2018neuro,yao2017face,wang2014ferroelectric}. Different variants of transistors and selector devices are used in the literature for the crossbar architecture for improving the crossbar performance. Architecture based in 1T1M synapses allows to remove the leakages which cause the reduction of output current in read transistors. However, this architecture of the synapses has significantly larger on-chip area and power consumption, comparing to single memristor (1M) crossbar architectures. In this paper, we avoid the application of $1T1M$ synapses and investigate the application of $1M$ synapses to maintain small on-chip area and low power consumption, and use the modular approach to reduce the leakage current problem and make the programming of the memristive arrays less complicated. 
As illustrated in \cite{dasha}, modular approach allows to reduce the leakage currents in the crossbar.
In this approach, the large crossbar is divided into smaller crossbars as shown in Fig. \ref{mod}, and the current from all modular crossbars is summed up to process through the activation function in MB1. As illustrated in simulation results, this approach allow to achieve similar performance accuracy, as single crossbar approach. 
}

{
In addition, if the network is scaled, the sequential processing can introduce the limitation to the system in the form of reduced processing speed. In this case, the parallel processing can be introduced, which involves the concurrent computation and simultaneous execution of the output computations. The modular approach can be useful as well, however, each modular crossbar should have corresponding processing blocks for backpropagation algorithm (MB1, MB2, MB3 and MB4). This introduces additional complexity for the system and increases on-chip area and power but reduces the processing time. 
Modular approach may also allow to remove the analog storage units. As the size of the crossbar will be reduced, a time delayed signal produced by a signal delay circuit can be used instead of analog storage unit. 
}

\section{Learning architectures}
\label{s4}

\begin{figure*}
\centering
\includegraphics[width=180mm]{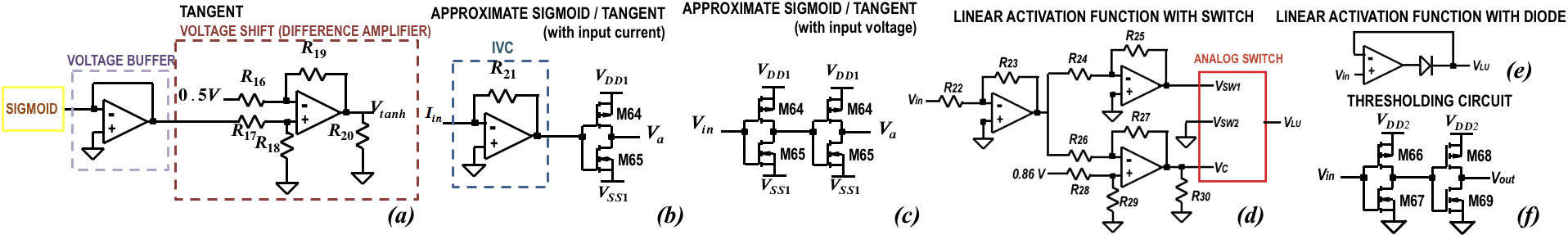}
\caption{{Implementation of various activation functions for DNN: (a) Tangent function based on the sigmoid circuit from Fig. \ref{cnew} (b). This architecture allows to implement a single circuit for both sigmoid and tangent functions in a multilayer neural network or another learning architecture and switch between these two functions. (b) Approximate sigmoid and approximate tangent functions driven by input current. To implement the sigmoid and tangent, the voltage levels $V_{DD1}$ and $V_{SS1}$ are varied. (c) Approximate sigmoid and approximate tangent functions driven by input voltage. (d) Linear activation function based on analog switch. (e) Linear activation function based on diode. (f) Additional thresholding circuit to normalize the neural network output for binary outputs. }}
\label{f9}
\end{figure*}

The proposed analog memristive backpropagation learning circuits can be used for various applications and learning architectures, such as neural networks, HTM and LSTM hardware implementations. To apply the proposed backpropagation circuits for various architectures, the implementation of additional functional blocks and activation functions is required. { In this section, we illustrate the implementations of tangent, current and voltage driven approximate sigmoid and tangent, and linear activation functions and thresholding circuit to normalize the output of the neural network.}

To implement the tangent function, the sigmoid function can be adjusted. The use of the sigmoid circuit (Fig. 3(b)) allows building a single circuit for both of the functions and switch between the sigmoid and tangent implementations when it is required. The implementation of the tangent function is shown in Fig. \ref{f9}(a). The sigmoid and buffer part remain the same as in the sigmoid implementation and the voltage shift circuit based on the difference amplifier is added. The difference amplifier is based on the same OpAmp shown in Fig. \ref{cnew}(c) with $R_{16}=10k\Omega$, $R_{17}=1k\Omega$, $R_{18}=2.5k\Omega$, $R_{19}=15k\Omega$ and $R_{20}=1k\Omega$. The circuit shifts the voltage level of the sigmoid and allows to implement a tangent function with the same circuit.

The implementation of an approximate sigmoid and tangent functions can be done with a simple thresholding circuit shown in Fig.\ref{f9}(b) and Fig.\ref{f9}(c). There are options: current-control and voltage-control approximate functions. The current-control circuit is shown in Fig. \ref{f9}(b). The input current $I_{in}$ is applied to the current-to-voltage converter based on the OpAmp with $R_{22}=20k\Omega$ and inverted by the inverter with $M64=0.18\mu/0.36\mu$ and $M65=0.18\mu/1.72\mu$. The $W/L$ ratio of $M64$ and $M65$ can be adjusted depending on the required transition part between the high and low value of approximate sigmoid and tangent functions. The voltages $V_{DD1}$ and $V_{SS1}$ are different for sigmoid and tangent implementations. For the approximate sigmoid $V_{DD1}=1V$ and $V_{SS1}=0V$, which means that the transistors have an under-drive voltage level for TSMS 180nm CMOS technology. In the approximate tangent implementation, $V_{DD1}=1V$ and $V_{SS1}=-1V$. The simple thresholding circuits can implement the voltage-controlled sigmoid and tangent with two inverters shown in Fig. \ref{f9}(c). The $W/L$ transistor ratios of $W_{66}-W_{69}$ and voltage levels of $V_{DD1}$ and $V_{SS1}$ can be adjusted to obtain a required amplitude, range and transition region for the sigmoid and tangent.

{
The implementation of linear activation functions are shown in Fig. \ref{f9}(d) and Fig. \ref{f9}(e). Both units are driven by voltage and the OpAmp circuits are shown in Fig. \ref{cnew}(c). We propose the implementation of linear activation function based on analog switch shown in Fig. \ref{f9}(d). The analog switch selects the $0V$ output for negative input signal, and $V_{in}$ for positive input. The switch is controlled by the shifted inverted input signal which is fed to the switch control $V_c$. All values of resistances are set to $1k\Omega$. The possibility to implement linear activation function is to use diode (Fig. \ref{f9}(e)). The diode based linear activation function has smaller on-chip area and lower power consumption, however the output range is smaller comparing to linear activation function with switch. To implement a current controlled linear activation function, IVC can be used in both circuits. In linear activation unit based on the analog switch, OpAmp $R_{22}-R_{23}$ can be replaced by IVC. In diode based linear activation function. additional IVC component before OpAmp is required.
}

{
As we verified with the simulation results, if the ideal outputs is required to  be binary, additional thresholding circuit is required in the output layer to normalize the outputs and achieve high accuracy. This was demonstrated using XOR problem in simulation results. The thresholding circuit that has been used for simulations is shown in  Fig. \ref{f9}(f), where the parameters of the transistors are $M66=M68=0.36\mu /0.18\mu$ and $M67=M69=0.72\mu /0.18\mu$. The thresholding function is connected to the output layer after the training process and allows to increase the accuracy significantly during the testing stage.
}

\subsection{Neural Networks}

There is a number of neural network architectures, where the proposed learning circuit can be used. The complete analog learning and training circuits for most of the architectures and networks covered in Section \ref{s4} has not been implemented in analog hardware. There are different architectures and types of the neural networks that can use the proposed learning circuit without making a significant modification to the proposed design, such as DNN, BNN, and MNN.

\subsubsection{Deep Neural Network}

\begin{figure}
\centering
\includegraphics[width=90mm]{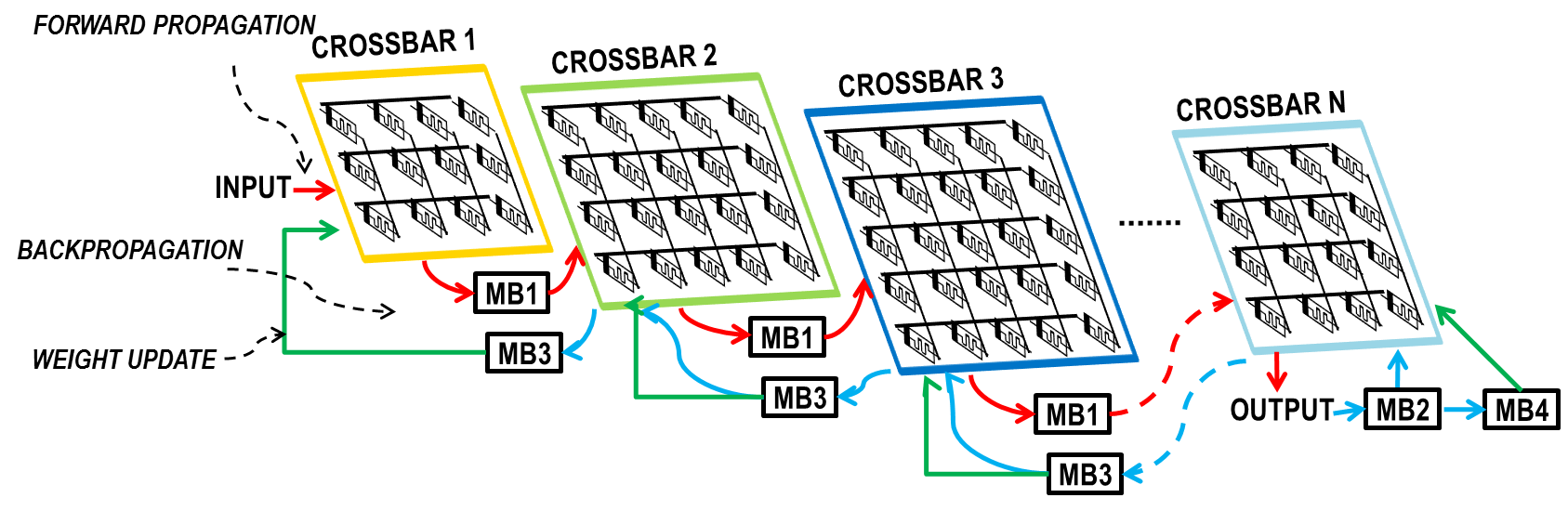}
\caption{Deep neural network implementation with the backpropagation learning. Red arrows correspond to forward propagation process. black arrows refer to backpropagation process. Green arrows show the weight update process.}
\label{f11}
\end{figure}

The proposed configuration for DNN with memristive analog learning circuits is shown in Fig. \ref{f11}. The DNN configuration contains N+1 layers, and N crossbars correspond to the synapses between the layers. In the forward propagation process, MB1 is used. MB1 can be modified to implement various activation functions. MB2 and MB4 perform the backpropagation process through the output layer of DNN, and MB3 does the backpropagation through hidden layers. In the update process, MB4 and MB3 are applied. The blocks MB2, MB3 and MB4 in each layer can be modified depending on the activation function applied in forward propagation in MB1 for each layer.

\subsubsection{Binary Neural Network}

\begin{figure}
\centering
\includegraphics[width=70mm]{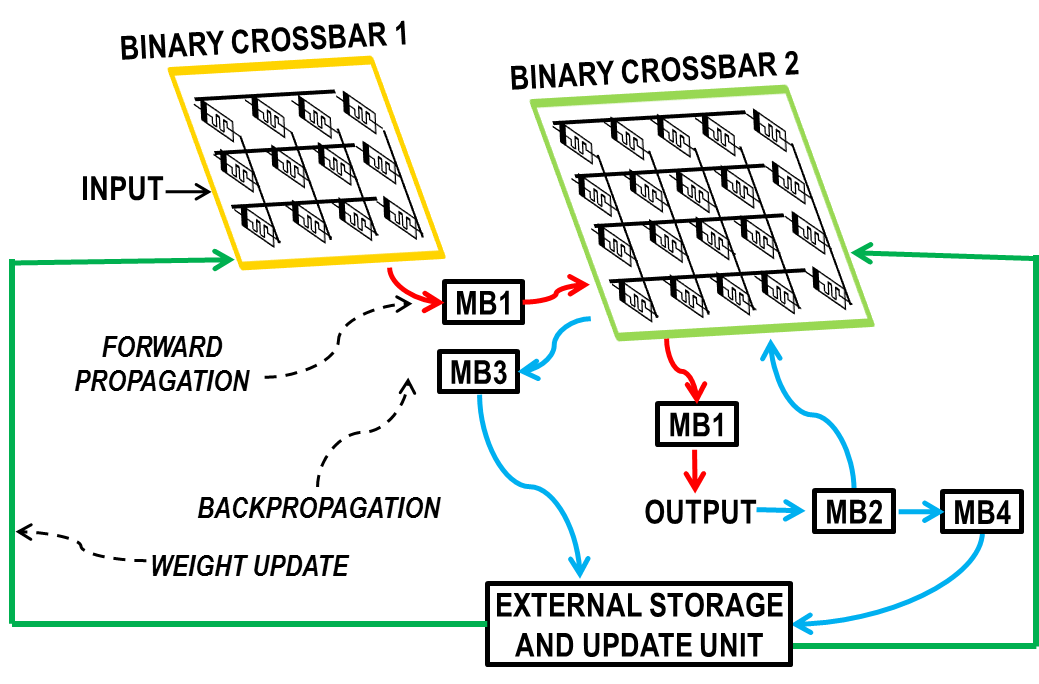}
\caption{Three layer BNN with backpropagation learning. The crossbars contain memristors that can be programmed only for $R_{ON}$ and $R_{OFF}$ stages. To improve the accuracy and the performance of the network, the change in error is stored in the external storage and update unit. The crossbar weights are updated based on several iterations in time.}
\label{f12}
\end{figure}

BNN can be implemented with the proposed circuit using two-stage memristors in the memristive crossbar representing the weights. The implementation of the BNN is shown in Fig. \ref{f12}. In BNN, the forward propagation process and backpropagation process are the same as in a three-layer neural network shown in Section \ref{s3}. However, due to the limitations of the binary weights, the direct update of the weights after the error calculation will not provide high accuracy results. We suggest to store the value of the change in error in the external storage and training units in time and, after a certain period of the training, update the weights. This method can improve the accuracy results for the classification problems using BNNs. 

\subsubsection{Multiple Neural Network}

\begin{figure}
\centering
\includegraphics[width=80mm]{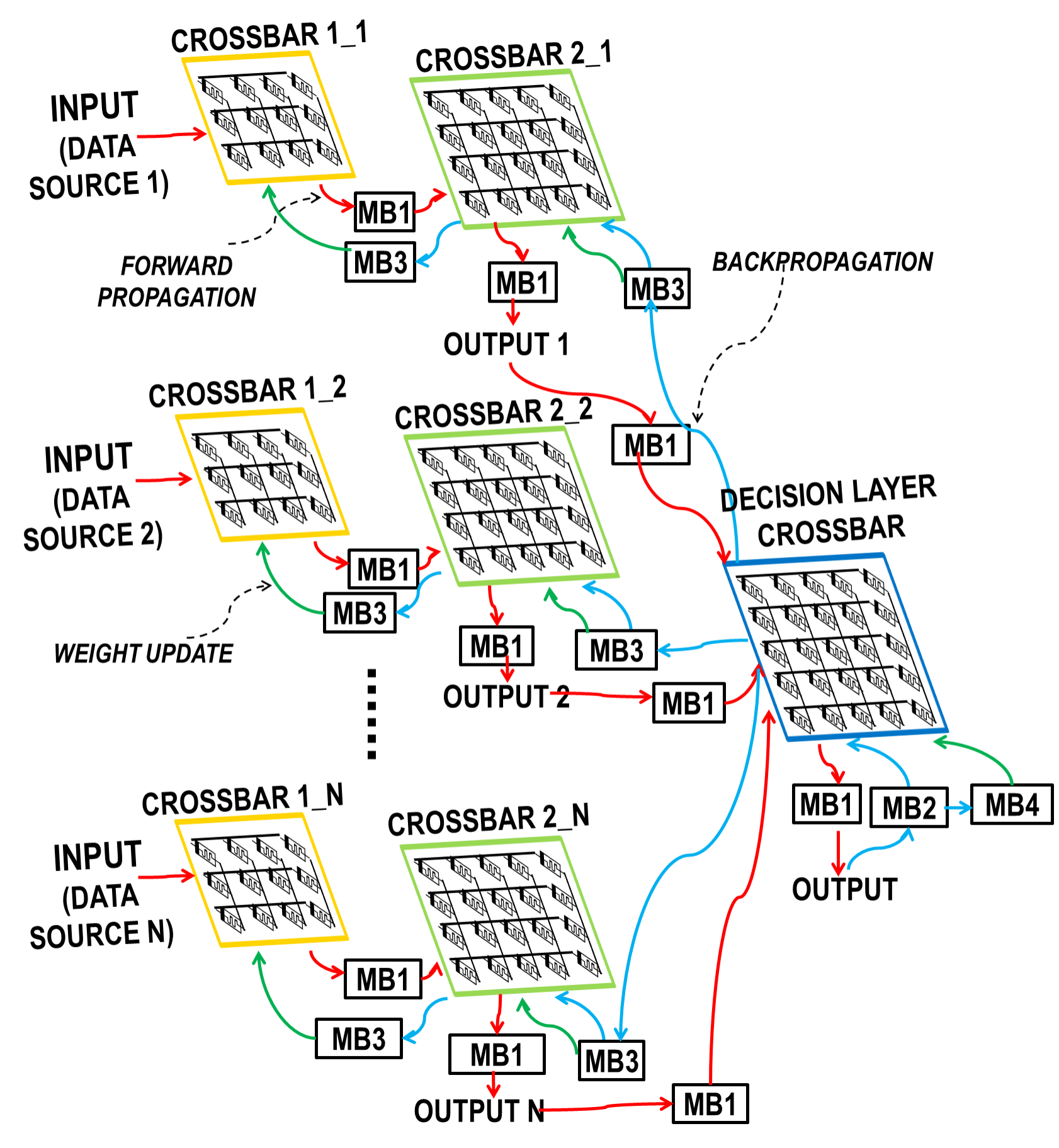}
\caption{Multiple neural network with backpropagation learning. The inputs from different data sources are fetched into different crossbars and the outputs from the crossbars are used as the inputs to the decision layer containing the memristive synapses.}
\label{f13}
\end{figure}

The proposed memristive analog learning circuits can be used for the MNN approach. This is useful when several sources of input data are used, and the decision on the output depends on the fusion of the results from each data source. The output from the different data sources are normalized, and outputs of each data source are fetched to separate crossbars, and the outputs of all the crossbars are fed into the decision layer. The decision layer crossbar is the same as the other crossbars in the system. As MNN consists of several networks and the decision layer can be treated as a separate network, all the networks can be trained either separately \cite{363444} or as a single network. The architecture of MNN with backpropagation learning that is trained as a single network is illustrated in Fig. \ref{f13}. We recommend using this approach, when the neural network inputs are taken from different data sources, such as various sensors in the system. One of the examples of the use of such system is gender recognition, where voice signals and face images can be used as inputs to MNN. The activation functions of the layer are different for separate crossbars and the decision layer and depend on the data that is used for the processing. The number of the required backpropagation blocks is equaled to the number of the crossbars that a network contains.

\subsection{Hierarchical Temporal Memory}

The other learning architecture, where the proposed circuit can be used is HTM. There are several hardware implementations proposed for the HTM SP and HTM TM \cite{ibrayev2017chip,fedorova2016htm,tcad}. However, the learning stage of the HTM SP has not been implemented on hardware yet. This stage can include either update process based on Hebb`s rule or the backpropagation update of the HTM SP weights \cite{25}. 

\begin{figure}
\centering
\subfigure[]
{
\includegraphics[width=87mm]{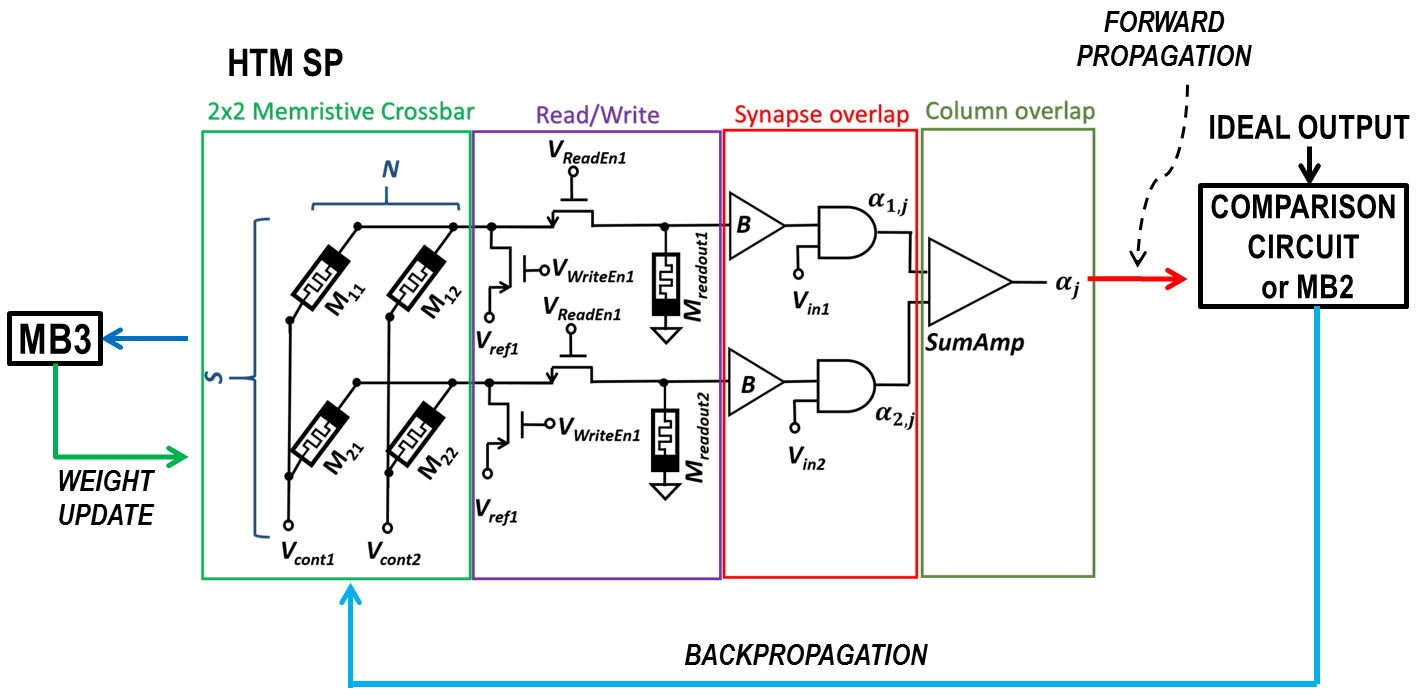}
\label{f14a}
}
\subfigure[]
{
\includegraphics[width=87mm]{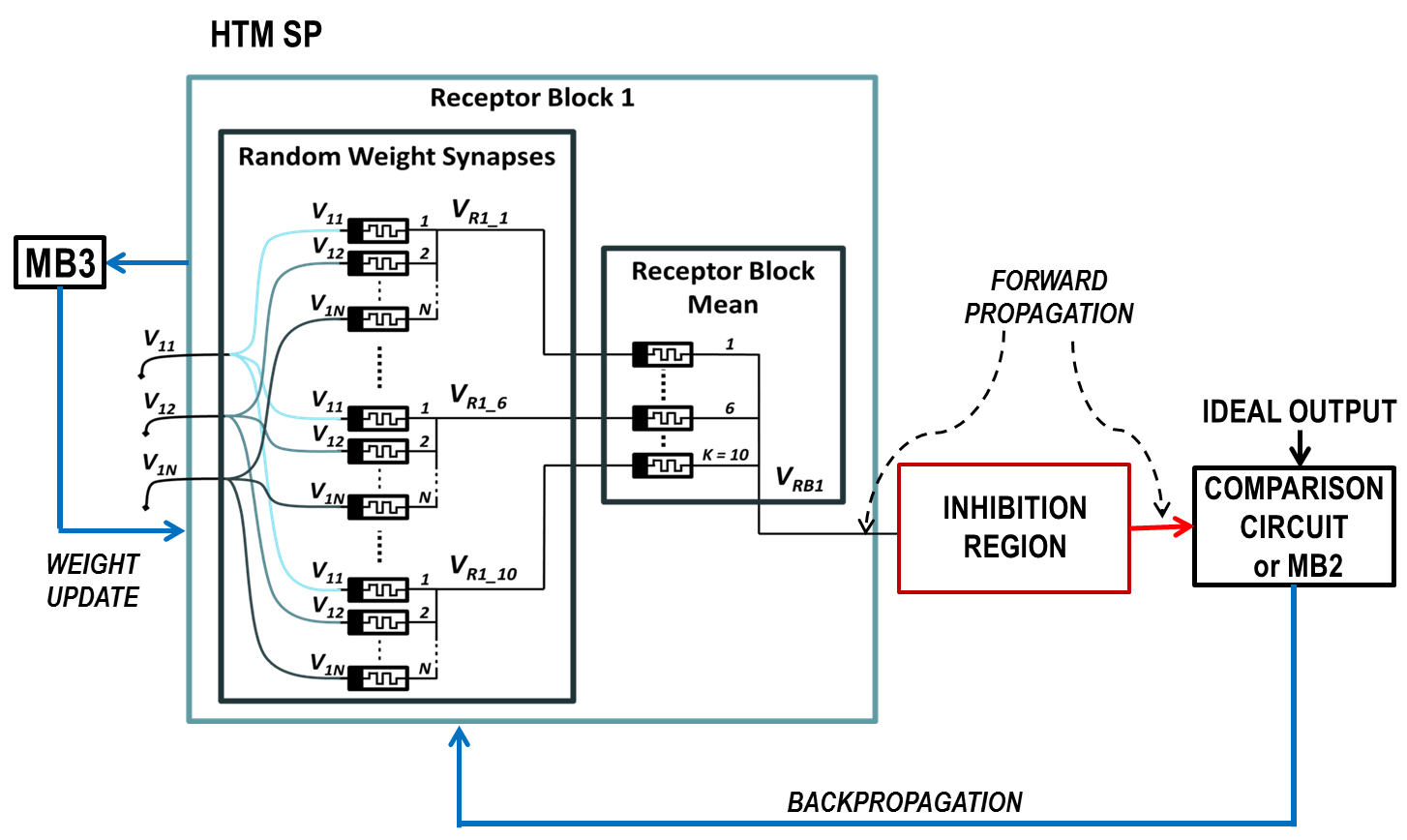}
\label{f14b}
}
\caption{Analog hardware implementation of the (a) conventional HTM SP algorithm and (b) modified HTM SP algorithm with backpropagation learning stage.}
\label{f14}
\end{figure}

There are two main analog architectures for the HTM SP: conventional HTM SP \cite{fedorova2016htm} and modified HTM SP \cite{tcad}. The application of the proposed learning circuits for both architectures is shown in Fig. \ref{f14}. Fig. \ref{f14a} illustrates the application of the proposed learning architecture for the conventional HTM circuits. After the forward propagation through the HTM SP, the HTM SP output is compared to the ideal HTM output. In the conventional HTM SP circuit, the calculated error from the comparison circuit or MB2 is fetched back to the memristive crossbar to calculate the error in the weights, and the weights are updated. Fig. \ref{f14b} shows the application of the proposed circuits for the modified HTM SP architecture. The conventional HTM SP architecture consists of the receptor and inhibition blocks. The weights of the synapses are located in the receptor block. After the comparison of the HTM SP output to the ideal output, the error is propagated back through the receptor block and MB3, and the memristive weights are updated.

\subsection{Long Short Term Memory}

\begin{figure}
\centering
\includegraphics[width=87mm]{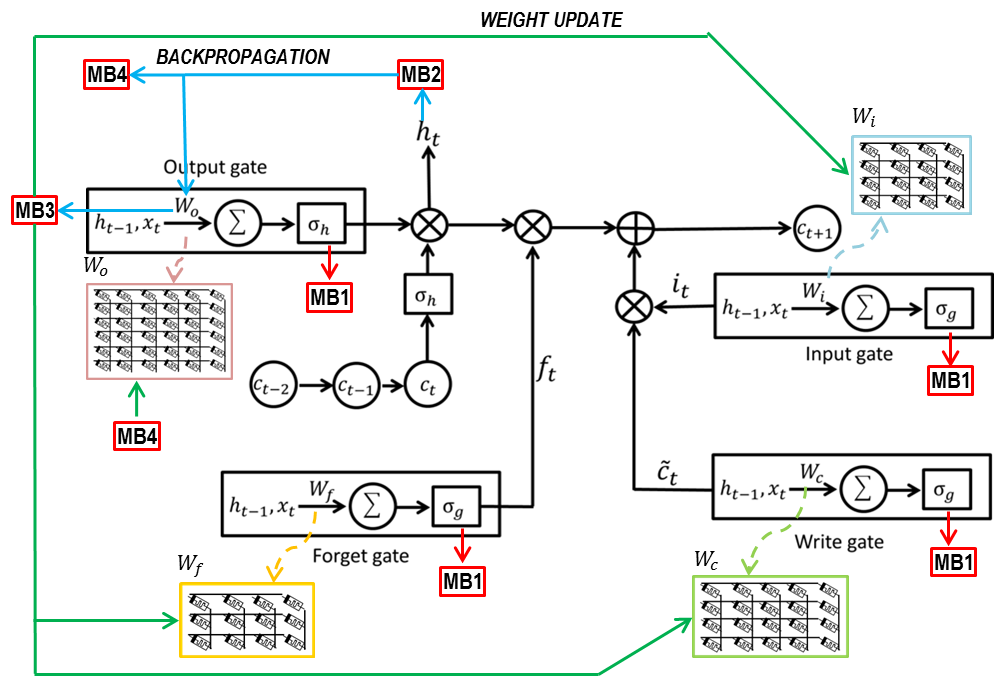}
\caption{Analog memristive hardware implementation of the LSTM algorithm.}
\label{f15}
\end{figure}

LSTM architecture can also be implemented using the proposed memristive analog backpropagation circuits. The full implementation of LSTM with analog circuits has not been proposed yet. However, the analog implementation of the separate LSTM components has been shown in \cite{kam,smagulova2018design}. Fig. \ref{f15} illustrates the full system level LSTM architecture consisting of the output gate, input gate, write gate and forget gate. The weights of LSTM gates $W_i$, $W_o$, $W_f$ and $W_c$ can be stored in the memristive crossbars. The activation functions in the LSTM architecture can be replaced with different variations of MB1. While, the weight update process of the crossbar $W_o$ is performed by MB4 as the update of the output layer, while MB3 performs the update of $W_i$, $W_f$ and $W_c$.

\section{Simulation results}
\label{s5}

The circuit simulations were performed in SPICE, and the verification of the ideal backpropagation algorithm is done in MATLAB. 

\subsection{Circuit performance}

\begin{figure}[!t]
    \centering        
    \subfigure[]
    	{
    	\includegraphics[width=80mm]{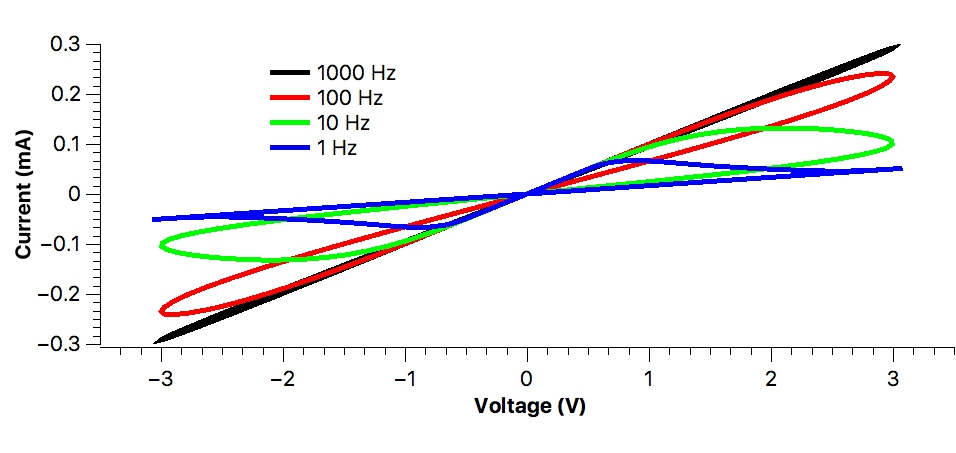}
		}       
     \subfigure[]
		{
    	\includegraphics[width=80mm]{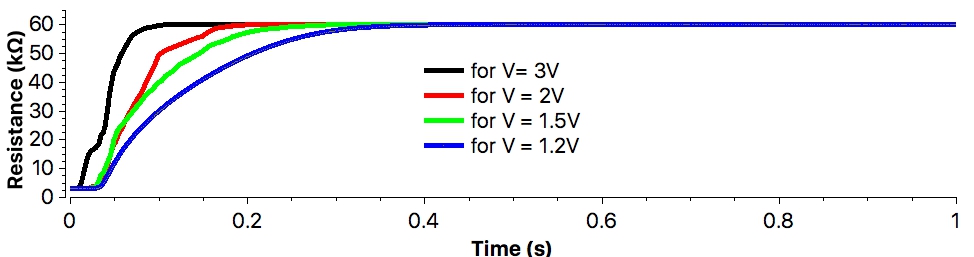}
		}        
     \subfigure[]
		{
    	\includegraphics[width=80mm]{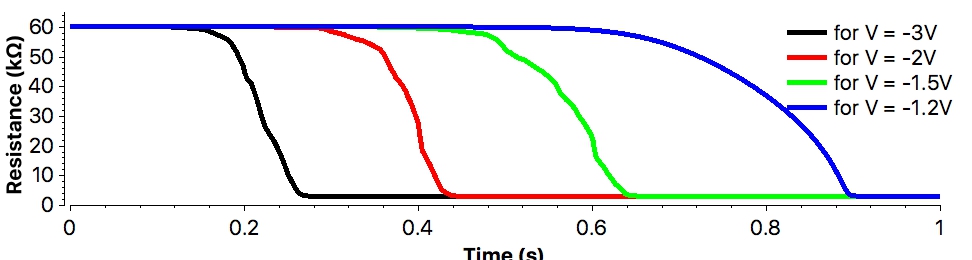}
		}
    \caption{Memristor characteristics: (a) hysteresis for different frequencies for $R_{initial}=10k\Omega$, (b) switching time from $R_{ON}=3k\Omega$ to $R_{OFF}=62k\Omega$ for different applied voltage amplitudes, and (c) switching time from $R_{OFF}=62k\Omega$ to $R_{ON}=3k\Omega$ for different applied voltage amplitudes.} 
    \label{memristor}
\end{figure}

The memristor model used in the crossbar simulations is Biolek`s modified S-model \cite{7527252} for HP $\textrm{TiO}_2$ memristor with the threshold voltage  $V_{th}$ of 1V \cite{strukov2008missing}. This memristor model is developed for large-scale simulations to simplify the computation and processing \cite{7527252}. The memristor characteristics and switching time for $R_{ON}=3k\Omega$ and $R_{OFF}=62k\Omega$ are shown in Fig. \ref{memristor}. {Fig. \ref{memristor} (b) and Fig. \ref{memristor} (c) illustrate memristor updated process applying pulse of 1s with different amplitudes.} The switching time is large, and speed of learning process with the memristive elements is slow. However, the learning and training process is a one-time process in the neural network. After the training during the testing stage, the reading time is small, and the data processing is fast. 

\begin{figure*}[!ht]
    \centering        
    \includegraphics[width=180mm]{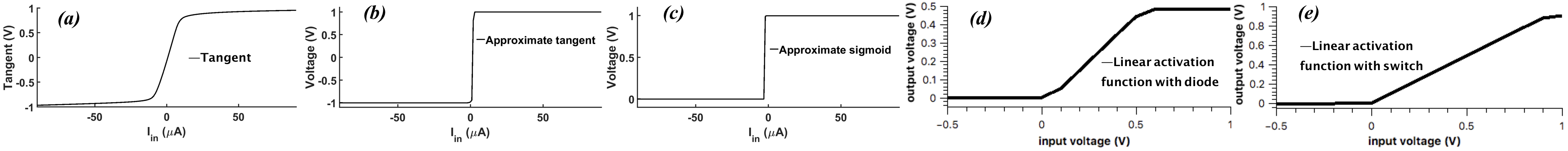}       
         \caption{{Simulation of additional activation functions versus current: (a) tangent, (b) approximate current driven sigmoid and (c) approximate current driven tangent, (d) linear activation with diode, (e) linear activation with switch.}}
    \label{add1}
\end{figure*}

\begin{figure}[!ht]
    \centering        
    \includegraphics[width=90mm]{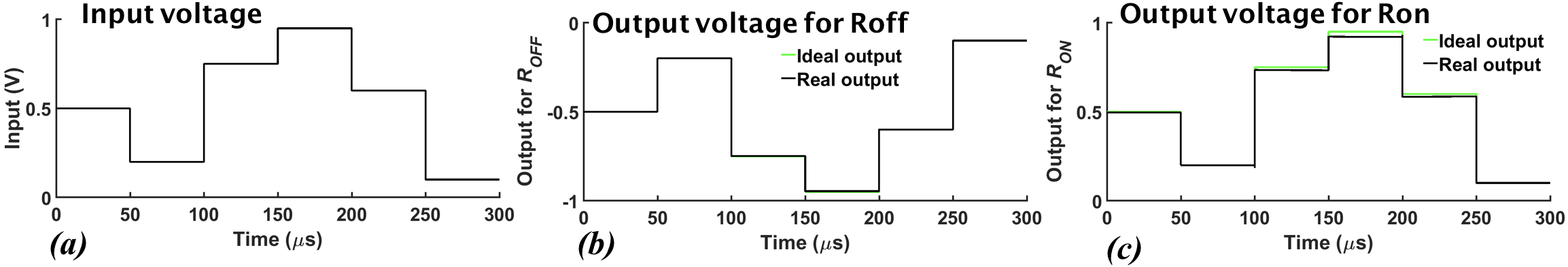}     
         \caption{Timing diagram for memristive weight sign control  circuit implementation: (a) input to the circuit, (b) ideal output and real output for $R_{ON}$ (when the weight is positive) and (c) ideal and real outputs for $R_{OFF}$ (when the weight is negative).}
    \label{add2}
\end{figure}

\begin{figure}[!ht]
    \centering        
    \includegraphics[width=90mm]{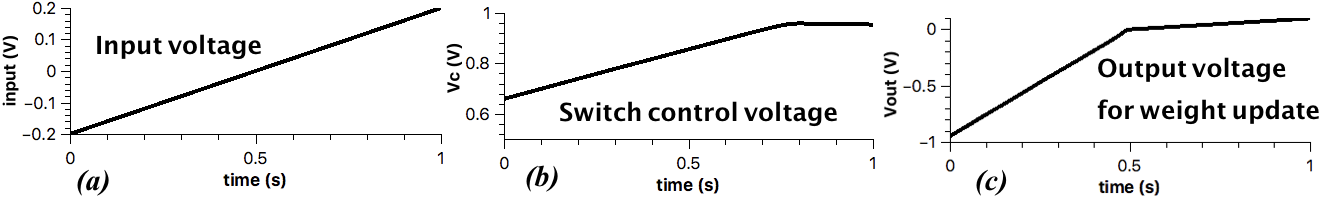}     
         \caption{{Output of weight update circuit: (a) input voltage, (b) switch control voltage, (c) output of weight update circuit programmed to high amplitude voltage for negative update voltages (update from $R_{OFF}$ to $R_{ON}$) and low amplitude voltage for positive update voltages ($R_{ON}$ to $R_{OFF}$).}}
    \label{add3}
\end{figure}

{
Simulation results illustrating the performance of the circuits in terms of amplitude are shown in Supplementary Material (Section IV).
}
The simulation results for additional activation functions are shown in Fig. \ref{add1}. Fig. \ref{add1} (a) represents the simulation of the proposed tangent function. Fig. \ref{add1} (b) and \ref{add1} (c) illustrate the simulation of current driven approximate sigmoid and tangent, respectively. 
{Fig. \ref{add1} (d) and \ref{add1} (e) show the simulation of linear activation functions with diode and switch, respectively. Fig. \ref{add1} shows that the activation function with switch is more linear. }
The timing diagram for the memristive weight sign control circuit is shown in Fig. \ref{add2}. Fig. \ref{add2} (a) represents the positive input voltage. Fig. \ref{add2} (b) illustrates the ideal output and real memristive weight sign control circuit output for $R_{ON}$, when the weight is positive. Fig. \ref{add2} (c) illustrates the ideal output and real output of the proposed weight sign control circuit for $R_{OFF}$, when the weight is negative.
{The output of weight update circuit is shown in Fig. \ref{add3}. The memristors is the circuit are programmed for high negative update amplitude and low positive update amplitude, as shown in Fig.  \ref{add3} (c).}

Table \ref{tableaa} represents the calculation of the on-chip area and maximum power dissipation for separate components for the analog learning circuit implementation and additional components and activation functions. Also, the example of the area and power dissipation for a small crossbar is shown.

\begin{table}[h]
{
\centering
\caption{Power consumption and on-chip area calculation for the separate circuit components.}
\label{tableaa}
\begin{tabular}{|p{3cm}|p{2cm}| p{2cm}|}
\hline
\textbf{Circuit component} & \textbf{Power consumption} & \textbf{On-chip area}                  \\ \hline
Crossbar (4 input neurons and 10 output neurons)                   & $5 \mu W $          & $1.36  \mu m^2 $ \\ \hline
Crossbar with control switches                   & $1200 \mu W $          & $ 115.3 \mu m^2 $ \\ \hline
Weight sign control circuit & $195.1 \mu W$& $16.64 \mu m^2$ \\ \hline
Sigmoid & $11.4 \mu W$& $184.00 \mu m^2$ \\ \hline
Current buffer             & $149.0 \mu W $         & $280.00 \mu m^2 $    \\ \hline
OpAmp (maximum)                      & $39.8 mW $           & $2801.76 \mu m^2 $ \\ \hline
Analog switch & $162.3 \mu W$ & $1.55\mu m^2 $\\ \hline
Approximate current driven sigmoid/tangent                     & $52.9 mW $           & $2118.00 \mu m^2 $ \\ \hline
Approximate voltage driven sigmoid/tangent                     & $ 41.2 pW$           & $0.40 \mu m^2 $ \\ \hline
Linear activation units with diode                     & $ 963.7 \mu W$           & $244 \mu m^2 $ \\ \hline
Linear activation unit with switch                     & $23.214 mW$           & $951.06 \mu m^2 $ \\ \hline
Weight update circuit                     & $ 14.34 mW$           & $1269.63 \mu m^2 $ \\ \hline
\end{tabular}
}
\end{table}

Table \ref{table11} shows the on-chip area and maximum power dissipation for separate components for the main blocks of the proposed analog backpropagation learning circuit implementation.

\begin{table}[h]
\centering
\caption{Area and power calculations for the main blocks of the proposed design and total area and power for three layer network.}
\label{table11}
\begin{tabular}{|p{3.5cm}|p{1.7cm}|p{1.7cm}|}
\hline
{Configuration}& {Area ($\mu m^2)$}  & {Maximum Power($mW$)} \\
\hline
\hline
MB1 (hidden layer)  & 4885.86  & 3.70 \\
\hline
MB2 + MB1 (output layer) & 8264.88 &  10.64   \\
\hline
  MB3 & 15238.69  & 61.78  \\
   \hline
  MB4 &  9734.33  & 39.53  \\
\hline
\hline
\textbf{Total} &  \textbf{38123.76}  & \textbf{115.65} \\
\hline
\end{tabular}
\end{table}


\subsection{System level simulations}
\begin{figure}[!ht]
    \centering        
    \includegraphics[width=70mm]{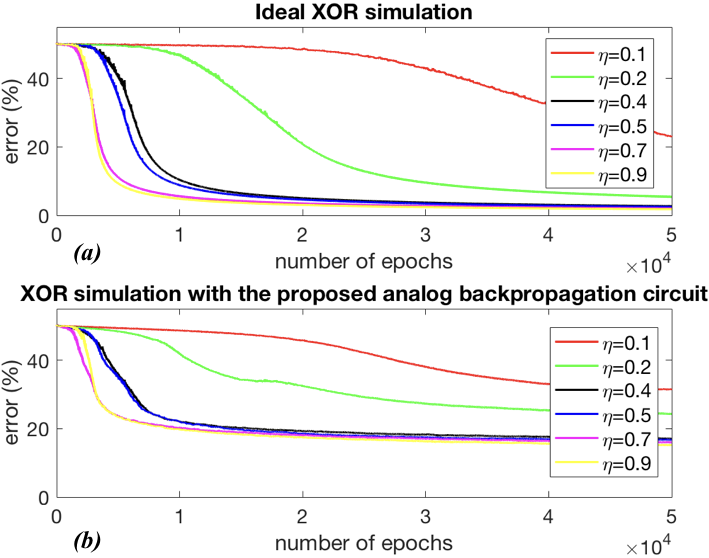}   
         \caption{{Error rate versus number of iterations for (a) simulations ideal algorithm and  (b) simulation of proposed backpropagation circuit.}}
    \label{xor}
\end{figure}

The system level simulations have been performed for XOR problem and handwritten digits recognition for ANN and face recognition for DNN. For setup in XOR problem, there are 2 neurons in the input layers, 4 neurons in the hidden layer and 1 neuron in the output layer. {During the training, input was selected randomly out of four possible inputs. } The simulation results for XOR problem for different learning rates for ideal simulations and backpropagation circuit are shown in Fig. \ref{xor} for number of iterations $n=50,000$. In the real circuit with non-ideal behavior, the error after $50,000$ iterations is 15\% higher than in ideal simulations. We verified that it is caused by the non-ideal behavior of analog multiplication circuits, which will be improved further in the future work. The accuracy results for XOR simulation for the cases with and without thresholding circuits are shown in Table III.

\begin{figure}[!ht]
    \centering        
    \includegraphics[width=90mm]{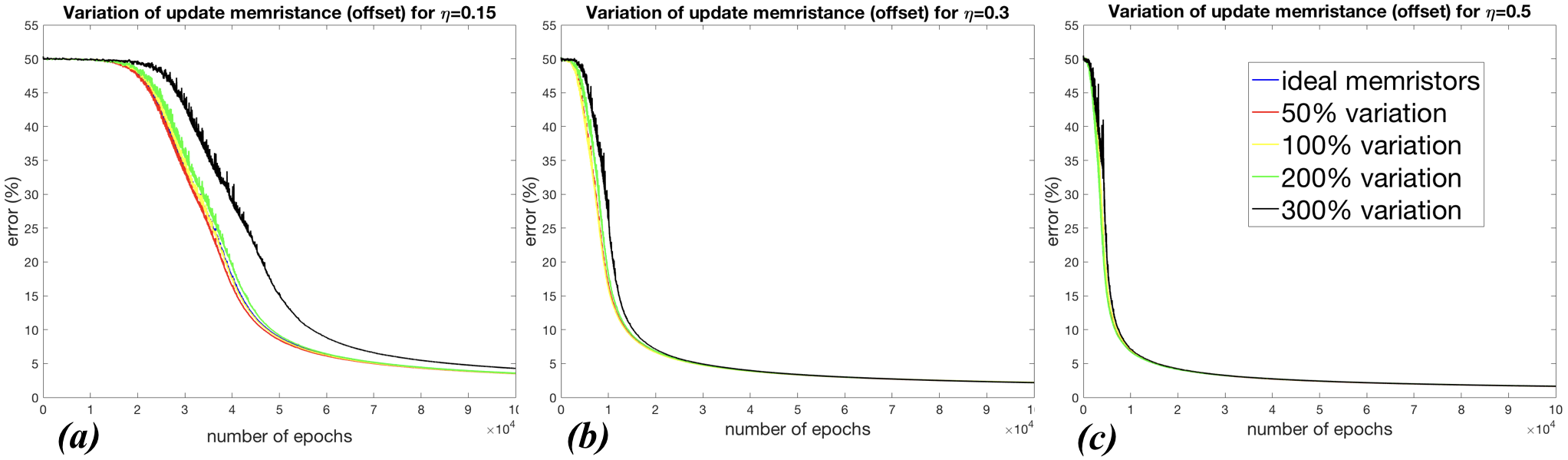}   
         \caption{{ Effect of the offset of the memristor update value on the performance of the architecture for (a) $\eta=0.15$,  (b)  $\eta=0.3$ and (c)  $\eta=0.5$ .}}
    \label{xor1}
\end{figure}

{
The variability analysis  for random offsets in memristor programming value is shown in Fig. \ref{xor1} for learning rates $\eta=0.15$,  $\eta=0.3$ and  $\eta=0.5$. The offset in the weight update value is represented as following: $w=w+(\Delta w + \Delta w \cdot x)$, where $x$ is a  random variation of the weight of particular percentage. This variation can be caused by non-ideal behavior of processing, update circuit and control circuits for update pulse duration. The architecture was tested for the variation of 50\%, 100\%, 200\% and 300 \%. The simulation results on Fig. \ref{xor1} show that the variation in the update value does not have a significant effect on the performance of the architecture. Also, the offsets in update value affect the system with smaller learning rate ($\eta=0.15$) more than the system with larger learning rate ($\eta=0.5$). However, the value of error converges to small error in all the cases. Therefore, even the significant error in the memristor update value, does not effect the performance of the learning process. The final accuracy for 100,000 iterations is illustrated in Table III. 
}

\begin{figure}[!ht]
    \centering        
    \includegraphics[width=90mm]{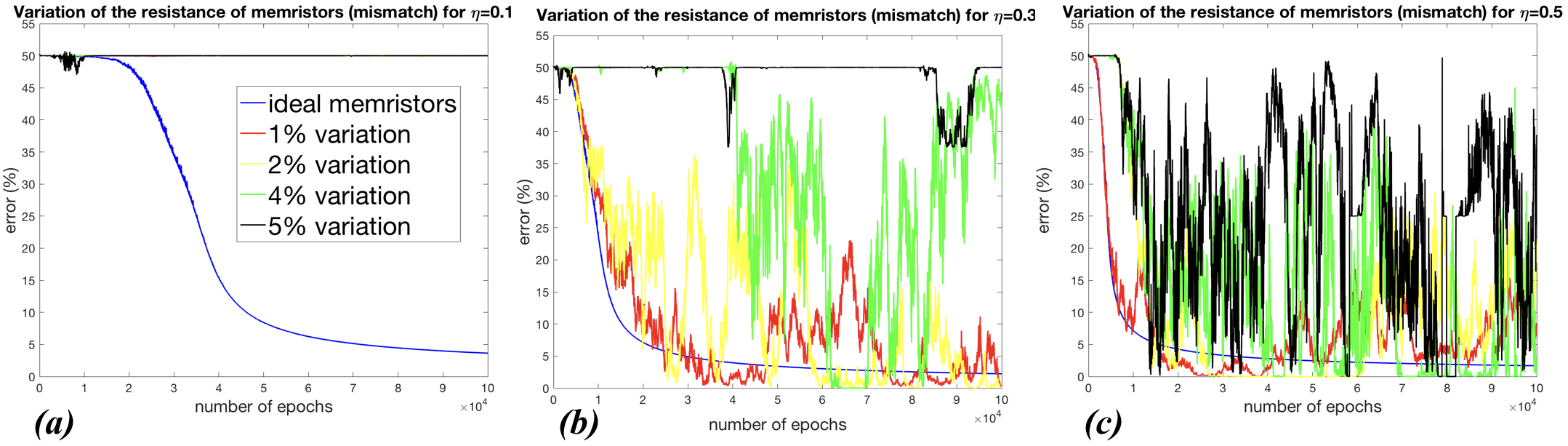}   
         \caption{{ Mismatch in the final weight of memristor of 1\%, 2\%, 4\% and 5\% for (a) $\eta=0.15$  (b) $\eta=0.3$ (c) $\eta=0.5$.}}
    \label{xor2}
\end{figure}

{
The performance analysis for the case of random mismatch in final memristor values after update is shown Fig. \ref{xor2}. The performance accuracy after 100,000 iterations for all cases is shown in Table III.  The mismatch is defined as following: $w=(w+\Delta w)+ (w+ \Delta w )\cdot x$, where $x$ is the percentage of variation, shown in Fig. \ref{xor2} as 1\%, 2\%, 4\% and 5\%. This mismatch has more significant effect on the performance of the architecture. For the small learning rate (Fig. \ref{xor2} (a)), the mismatch in the memristor values does not allow system to converge. For the larger learning rates, the system converge slower that in ideal case for 1-2\% of mismatch and does not converge for larger mismatches. However, such case is the effect of the non-linear behavior and instability of memristive device, which should be investigated further at the device level. 
}

\begin{table}[]
\label{xoracc}
{
\caption{Accuracy for XOR simulations (2 input neurons, 4 hidden layer neurons and 1 output neuron).)}
\begin{tabular}{|l|l|c|c|}
\hline
\multicolumn{2}{|l|}{\textbf{\begin{tabular}[c]{@{}l@{}}Configuration\\  for XOR simulation\end{tabular}}}                                                                            & \textbf{\begin{tabular}[c]{@{}c@{}}ANN accuracy \\ (without \\ thresholding\\ circuit)\end{tabular}} & \textbf{\begin{tabular}[c]{@{}c@{}}ANN accuracy\\  (with \\ thresholding\\ circuit, $\theta=0.5$)\end{tabular}} \\ \hline \hline
\multicolumn{2}{|l|}{\begin{tabular}[c]{@{}l@{}}Ideal memristors\\  $\eta=0.15$, $n=50,000$\end{tabular}}                                                                             & 84.8\%                                                                                               & 100\%                                                                                                           \\ \hline
\multicolumn{2}{|l|}{\begin{tabular}[c]{@{}l@{}}Ideal memristors\\  $\eta=0.15$, $n=100,000$\end{tabular}}                                                                            & 96.26\%                                                                                              & 100\%                                                                                                           \\ \hline
\multicolumn{2}{|l|}{\begin{tabular}[c]{@{}l@{}}Ideal memristors\\  $\eta=0.3$, $n=100,000$\end{tabular}}                                                                             & 97.76\%                                                                                              & 100\%                                                                                                           \\ \hline
\multicolumn{2}{|l|}{\begin{tabular}[c]{@{}l@{}}Ideal memristors\\  $\eta=0.5$, $n=100,000$\end{tabular}}                                                                             & 98.31\%                                                                                              & 100\%                                  \\ \hline
\hline
\begin{tabular}[c]{@{}l@{}}Offset in memristors\\ programming value\\ $\eta=0.15$, \\ $n=100,000$\end{tabular} & \begin{tabular}[c]{@{}l@{}}50\%\\ 100\%\\ 200\%\\ 300\%\end{tabular} & \begin{tabular}[c]{@{}c@{}}96.10\%\\ 96.10\%\\ 96.07\%\\ 96.10\%\end{tabular}                        & \begin{tabular}[c]{@{}c@{}}100\%\\ 100\%\\ 100\%\\ 100\%\end{tabular}                                           \\ \hline
\begin{tabular}[c]{@{}l@{}}Offset in memristors\\ programming value\\ $\eta=0.3$,\\ $n=100,000$\end{tabular}   & \begin{tabular}[c]{@{}l@{}}50\%\\ 100\%\\ 200\%\\ 300\%\end{tabular} & \begin{tabular}[c]{@{}c@{}}96.48\%\\ 96.48\%\\ 96.41\%\\ 95.72\%\end{tabular}                        & \begin{tabular}[c]{@{}c@{}}100\%\\ 100\%\\ 100\%\\ 100\%\end{tabular}                                           \\ \hline
\begin{tabular}[c]{@{}l@{}}Offset in memristors\\ programming value\\ $\eta=0.5$,\\ $n=100,000$\end{tabular}   & \begin{tabular}[c]{@{}l@{}}50\%\\ 100\%\\ 200\%\\ 300\%\end{tabular} & \begin{tabular}[c]{@{}c@{}}98.56\%\\ 98.58\%\\ 98.56\%\\ 98.33\%\end{tabular}                        & \begin{tabular}[c]{@{}c@{}}100\%\\ 100\%\\ 100\%\\ 100\%\end{tabular}                                           \\ \hline
 \hline
\begin{tabular}[c]{@{}l@{}}Random mismatches\\ in memristor value\\ $\eta=0.15$,\\ $n=100,000$\end{tabular}    & \begin{tabular}[c]{@{}l@{}}1\%\\ 2\%\\ 4\%\\ 5\%\end{tabular}        & \begin{tabular}[c]{@{}c@{}}50.02\%\\ 50.82\%\\ 50\%\\ 50\%\end{tabular}                              & \begin{tabular}[c]{@{}c@{}}50\%\\ 50\%\\ 50\%\\ 50\%\end{tabular}                                               \\ \hline
\begin{tabular}[c]{@{}l@{}}Random mismatches\\ in memristor value\\ $\eta=0.3$,\\ $n=100,000$\end{tabular}     & \begin{tabular}[c]{@{}l@{}}1\%\\ 2\%\\ 4\%\\ 5\%\end{tabular}        & \begin{tabular}[c]{@{}c@{}}99.7\%\\ 99.89\%\\ 56.73\%\\ 50\%\end{tabular}                            & \begin{tabular}[c]{@{}c@{}}100\%\\ 100\%\\ 62.5\%\\ 50\%\end{tabular}                                           \\ \hline  
\begin{tabular}[c]{@{}l@{}}Random mismatches \\ in memristor value\\ $\eta=0.5$,\\ $n=100,000$\end{tabular}    & \begin{tabular}[c]{@{}l@{}}1\%\\ 2\%\\ 4\%\\ 5\%\end{tabular}        & \begin{tabular}[c]{@{}c@{}}91.77\%\\ 80.29\%\\ 99.22\%\\ 63.23\%\end{tabular}                        & \begin{tabular}[c]{@{}c@{}}100\%\\ 100\%\\ 100\%\\ 87.5\%\end{tabular}                                          \\ \hline 
\end{tabular}
 }
\end{table}

{ To verify the performance of the proposed approaches for real pattern recognition problems, we tested ANN for handwritten digits recognition and DNN for face recognition for 2 approaches: single crossbar (shown in Fig. \ref{f1}) and modular crossbar (shown in Fig. \ref{mod}) using $Ge_2Sb_2Te_5$ (GST) memristors with 16 resistive levels \cite{kuzum2011nanoelectronic,xiao2017gst}. 
In ANN simulation, MNIST database \cite{mnist} with $70,000$ images of the size of $28\times 28$ was used, where 86\% of images was selected for testing and 14\% for testing. The setup for ANN consisted of $28\times 28$ input layer neurons, 42 hidden layer neurons and 10 output neurons (corresponding to 10 classes of digits). In the modular approach, 16 crossbars with 49 input neurons, 8 crossbars with 98 input neurons and 4 crossbars with 196 input neurons were tested. For DNN verification, we performed face recognition task using Yale database for face recognition with 165 images of 15 people \cite{georghiades1997yale}. The images were rescaled by the size of $32\times 32$, and 45\% of the dataset was used for training and 55\% for testing. The DNN configuration consisted of 6 layers of 1024, 800, 500, 100, 30 and 15 neurons. The simulation results are shown in Table IV.  As the accuracy for all modular configurations is approximately the same, the modular crossbar approach is presented by a single value in the table. The simulation results show that the performance accuracy for both real ANN and DNN is reduced slightly, comparing to ideal case. As the obtained accuracy is the same and the research work \cite{dasha} shows that the leakage currents are reduced in modular approach, the crossbar with 1M synapses can be divided into modular crossbars to avoid 1T1M synapses and reduce the on-chip area of the crossbar.
}

\begin{table}[]
\label{anndnn}
{
\centering
\caption{ANN accuracy for handwritten digits recognition application and DNN accuracy for face recognition application.}
\begin{center}
\begin{tabular}{|l|c|c|}
\hline
\textbf{Configuration} & \textbf{\begin{tabular}[c]{@{}c@{}}ANN accuracy \\ (MNIST,\\  handwritten digits)\end{tabular}} & \textbf{\begin{tabular}[c]{@{}c@{}}DNN accuracy \\ (Yale, \\ face recognition)\end{tabular}} \\ \hline
Ideal simulations      & 93\%                                                                                            & 78.9\%                                                                                       \\ \hline
Single crossbar        & 92\%                                                                                            & 73.3\%                                                                                       \\ \hline
Modular crossbar       & 92\%                                                                                            & 75.5\%                                                                                       \\ \hline
\end{tabular}
\end{center}
}
\end{table}

\section{Discussion}
\label{s6}
The proposed analog hardware implementation of the backpropagation algorithm can be used to implement the online training of different learning architectures,  {which can be used for near-sensor processing.} The analog memristive learning architecture allows removing the additional software based or digital offline training and learning process. This can increase the processing speed and reduce the processing time, comparing to digital analogies, where the number of components to achieve high sampling rates in analog-to-digital  converters (ADC) and digital-to-analog converters (DAC) are large.  The possible errors in the training caused by leakage currents and parasitic effects can be mitigated by the increase of the number of iterations in the learning stage.

If the sneak path problems occur during the training state and the memristor update value is not accurate, this can be fixed in the following learning stages, but more update iterations are required for error to converge and reach high accuracy in this case.
As demonstrated in the XOR simulation, the problem of the non-ideal variation of the update value of the memristor due to non-ideal performance of the circuit and other device instabilities can be eliminated by increasing number of training iterations.

The limitations of the proposed architecture include the scalability of the memristive crossbar arrays and limitations of the current memristive devices. The problems of the parasitics, leakage current and sneak paths in the memristive crossbar have to be investigated further. The other drawback is a limitation of the memristive devices in terms of the number of resistance levels that can be achieved for particular memristive devices. 
The future work will include the implementation of the crossbar with a physically realizable memristor \cite{messaris2018data}, evaluation of the performance of the architecture with different memristive devices, evaluation of the abilities of different memristive devices and adjustment of circuit parameters for particular devices.

In addition, the limitations of the memristive devices, electromagnetic effects, frequency effects and their effect on the accuracy and the performance of the proposed learning architecture have to be studied. Also, the endurance of the memristive devices should be studied, especially for the case of several iterations in the learning process.


 The testing of the complete systems for large scale problems has to be performed and the limitations, such as loading effects and parasitics, have to be identified from the physical design constraints perspective. The effect of the additional components of the overall system performance and processing speed has to be determined under such conditions that become technology specific issues.
The future work will include the full circuit implementation of the proposed HTM, LSTM and MNN architectures and verification of their performance for large scale problems.


\section{Conclusion}
\label{s7}

In this paper, we presented the circuit design of an analog CMOS-memristive backpropagation learning circuit and its integration to different neural network architectures. The circuit architectures are presented for a three-layer neural network, DNN, BNN, MNN, conventional and modified HTM SP and LSTM. We presented the analog circuit implementation of interfacing circuits and activation functions that can be used to implement various learning architectures. The implementation of backpropagation with analog circuits offers simplicity of building differential operations combined with a dot-product operator as memristor crossbar that is useful of building neural networks. Using databases of MNIST (character recognition) and Yale (face recognition)  an application level validation of the proposed learning circuits for ANN and DNN architectures is successfully demonstrated.
The presented design of crossbar does not take into account physical design issues of memristive devices, while sneak path problem of crossbar arrays is accounted in the simulations by including conductance variability of real memristor devices and wire resistors in the crossbar.  However, the signal integrity issues is a topic to investigate further, when the memristor technology is mature and is suitable for a fabricating reliable large-scale arrays. The area and power of the proposed circuit design need to be further optimized a fully parallel implementation for real-time applications.

\ifCLASSOPTIONcaptionsoff
  \newpage
\fi

\bibliographystyle{IEEEtran}
\bibliography{reference}

\begin{IEEEbiography}[{\includegraphics[width=1in,height=1.25in,clip,keepaspectratio]{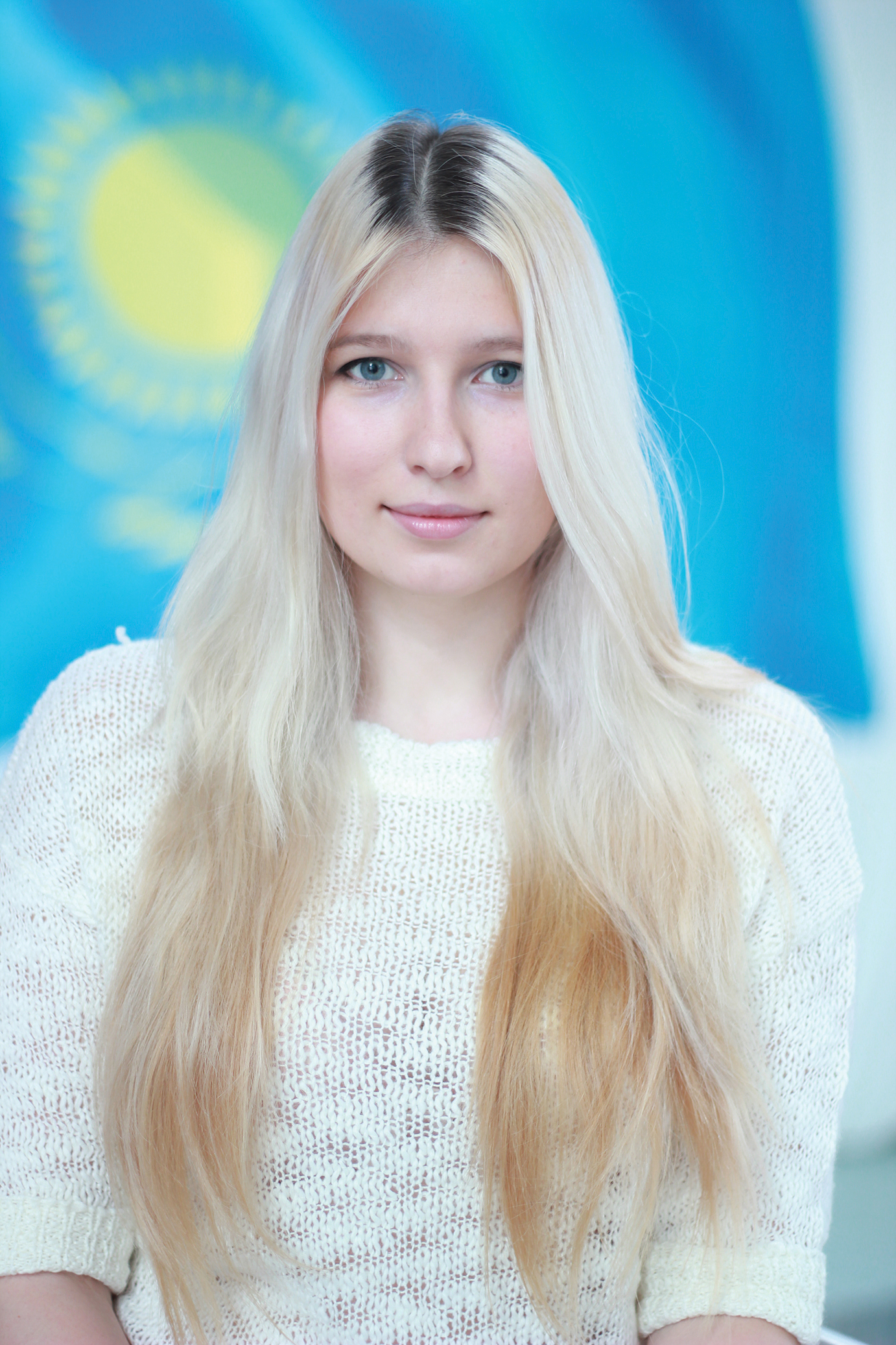}}]%
{Olga Krestinskaya}
  is working towards her graduate degree thesis in the area of neuromorphic memristive system from Electrical and Computer Engineering department at Nazarbayev University. She completed her bachelor of Engineering degree with honors in Electrical Engineering, with a focus on bio-inspired memory arrays in May 2016. Currently, she focuses on memristive circuits for hierarchical temporal memory, deep learning neural networks, and pattern recognition algorithms.  She is a Graduate Student Member of IEEE.
\end{IEEEbiography}

\begin{IEEEbiography}[{\includegraphics[width=1in,height=1.25in,clip,keepaspectratio]{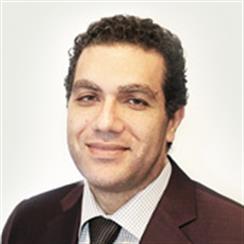}}]%
{Khaled N. Salama} (SM' 10) received the B.S. degree (Hons.) from the Department of Electronics and Communications, Cairo University, Cairo, Egypt, in 1997, and the M.S. and Ph.D. degrees from the Department of Electrical Engineering, Stanford University, Stanford, CA, USA, in 2000 and 2005, respectively.He was an Assistant Professor with the Rensselaer Polytechnic Institute, Troy, NY, USA, from 2005 to 2009. In 2009, he joined the King Abdullah University of Science and Technology, Saudi Arabia, where he is currently a Professor and was also the founding Program Chair until 2011. His work on CMOS sensors for molecular detection has been funded by the National Institutes of Health and the Defense Advanced Research Projects Agency, received the Stanford-Berkeley Innovators Challenge Award in biological sciences and was acquired by Lumina Inc. He has authored 225 papers and 14 patents on low-power mixed signal circuits for intelligent fully integrated sensors and nonlinear electronics, in particular memristor devices.
\end{IEEEbiography}

\begin{IEEEbiography}[{\includegraphics[width=1in,height=1.25in,clip,keepaspectratio]{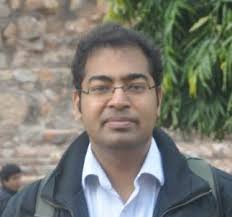}}]%
{Alex Pappachen James} (SM'13)
  works on brain-inspired circuits, memristor circuits, algorithms and systems, and has a PhD from Griffith School of Engineering, Griffith University.  He is currently the Vice Dean of research and graduate studies and also the Chair of Electrical and Computer Engineering department at Nazarbayev University. He is also the chair of IEEE Kazakhstan subsection.  He has a sustained experience of managing industry and academic projects in board design, VLSI and pattern recognition algorithms,  and semiconductor industry. He was editorial board member of Information fusion and is currently Associate Editor of Human-centric Computing and Information Sciences, IEEE Access, IEEE Transactions on Circuits and Systems 1, and served as guest associate editor to IEEE Transactions on Emerging Topics in Computational Intelligence. He is a Senior Member of IEEE and Senior Fellow of HEA. More see http://biomicrosystems.info/alex/
\end{IEEEbiography}

\vfill 
\end{document}